\documentclass[letterpaper]{JHEP3}
\usepackage{epsfig,multicol,bbm}

\newcommand{\roughly}[1]{\mathrel{\raise.3ex\hbox{$#1$\kern-0.85em
\lower1ex\hbox{$\sim$}}}}

\def\pd{\partial}
\def\proj{\mathcal P}

\def\cA{{\cal A}}

\def\cF{{\cal F}}

\def\cO{{\cal O}}

\def\cR{{\cal R}}

\def\cV{{\cal V}}

\def\cx{\mathcal{X}}
\def\cy{\mathcal{Y}}
\def\cz{\mathcal{Z}}

\def\exd{{\hbox{d}}}

\def\ba{\begin{eqnarray}}
\def\ea{\end{eqnarray}}
\def\be{\begin{equation}}
\def\ee{\end{equation}}

\def\ssM{{\scriptscriptstyle M}}
\def\ssN{{\scriptscriptstyle N}}
\def\ssP{{\scriptscriptstyle P}}
\def\ssQ{{\scriptscriptstyle Q}}
\def\ssR{{\scriptscriptstyle R}}
\def\ssT{{\scriptscriptstyle T}}
\def\ssV{{\scriptscriptstyle V}}
\def\ssW{{\scriptscriptstyle W}}
\def\KK{{\scriptscriptstyle KK}}
\def\GH{{\scriptscriptstyle GH}}
\def\EF{{\scriptscriptstyle EF}}
\def\JF{{\scriptscriptstyle JF}}

\def\nn{\nonumber}

\def\({\left(}
\def\){\right)}

\def\pref#1{(\ref{#1})}

\title{Sculpting the Extra Dimensions:\\
Inflation from Codimension-2 Brane Back-reaction}
\author{
Leo van Nierop${}^1$ and C.P. Burgess${}^{1,2}$ \\
$^1$Department of Physics and Astronomy, McMaster University, Hamilton ON, Canada\\
$^2$Perimeter Institute for Theoretical Physics, Waterloo ON, Canada\\
}

\maketitle

\abstract { We construct an inflationary model in 6D supergravity that is based on explicit time-dependent solutions to the full higher-dimensional field equations, back-reacting to the presence of a 4D inflaton rolling on a space-filling codimension-2 source brane. Fluxes in the bulk stabilize all moduli except the `breathing' modulus (that is generically present in higher-dimensional supergravities). Back-reaction to the inflaton roll causes the 4D Einstein-frame on-brane geometry to expand, $a(t) \propto t^p$, as well as exciting the breathing mode and causing the two off-brane dimensions to expand, $r(t) \propto t^q$. The model evades the general no-go theorems precluding 4D de Sitter solutions, since adjustments to the brane-localized inflaton potential allow the power $p$ to be dialed to be arbitrarily large, with the 4D geometry becoming de Sitter in the limit $p \to \infty$ (in which case $q = 0$). Slow-roll solutions give accelerated expansion with $p$ large but finite, and $q = \frac12$. Because the extra dimensions expand during inflation, the present-day 6D gravity scale can be much smaller than it was when primordial fluctuations were generated --- potentially allowing TeV gravity now to be consistent with the much higher gravity scale required at horizon-exit for observable primordial gravity waves. Because $p \gg q$, the 4 on-brane dimensions expand more quickly than the 2 off-brane ones, providing a framework for understanding why the observed four dimensions are presently so much larger than the internal two. If uplifted to a 10D framework with 4 dimensions stabilized, the 6D evolution described here could describe how two of the six extra dimensions evolve to become much larger than the others, as a consequence of the enormous expansion of the 4 large dimensions we can see. \\
 }

\begin{document}
\section{Introduction}

The spirit of most extra-dimensional models of particle physics is to translate observed or desirable properties of ordinary 4D particle interactions into particular shapes or features (like warping or brane positions) within an assumed extra-dimensional geometry. In principle these features are hoped to be obtained by minimizing the energy of deforming the extra dimensions, but it is in practice a challenge to do so explicitly.

Part of what makes this challenging is the fact that general covariance makes energy in itself not a useful criterion for distinguishing amongst various solutions. For instance for closed geometries invariance under time reparameterization implies {\em all} solutions have precisely zero energy. This has long been understood in cosmology, where the explanation of the geometry of the present-day universe is seen to be contingent on the history of how it evolved in the distant past. A similar understanding is also likely for the shapes of any present-day extra dimensions, suggesting we should seek to explain their properties in terms of how they have evolved over cosmological times.

This is not the approach taken by most models of extra-dimensional cosmology, however, which usually explicitly assume extra dimensions to be stabilized at fixed values as the observed four dimensions change in time. This approach is taken usually for technical reasons: it is difficult to find explicit time-dependent solutions to the full higher-dimensional field equations. Instead, models of extra-dimensional cosmology usually use one of two simplifying approximations: either `mirage' or `4D effective' cosmology.

In `mirage' cosmology \cite{MirageCosmology} brane-localized observers experience time-dependent geometries because they move through a static extra-dimensional bulk. In these models the branes are usually taken as `probe' branes, that don't back-react on the static bulk. An exception to this is for Randall-Sundrum type cosmologies \cite{RScosmo} involving codimension-1 branes, for which the Israel junction conditions \cite{IJC} allow back-reaction to be explicitly computed. In these models all extra-dimensional features are usually fixed from the get-go.

In `effective 4D' cosmology the Hubble scale is assumed to be much smaller than the Kaluza-Klein (KK) mass scale, so that all of the time dependence in the geometry can be computed within the effective 4D theory, where some extra-dimensional features (like moduli) boil down to the values of various scalar fields. This is the approach most frequently used for string inflation, for example \cite{SIreviews}. Here some changes to the extra dimensions can be followed by seeing how the corresponding modulus fields evolve. But this can only be done for sufficiently slow expansion and only after it is already assumed that the extra dimensions are so small that the 4D approximation is valid. In particular, it cannot follow evolution where all dimensions are initially roughly the same size, to explain why some dimensions are larger than others.

Our goal in this paper is to take some first steps towards going beyond these two types of approximations. To this end we explore the implications of previously constructed time-dependent solutions \cite{scaling solutions} to the full higher-dimensional field equations of chiral gauged 6D supergravity \cite{NS}, including the effects of back reaction from several codimension-2 source branes. When doing so it is crucial to work with a geometry with explicitly compactified extra dimensions, including a mechanism for stabilizing the extra-dimensional moduli, since it is well known that these can compete with (and sometimes ruin) what might otherwise appear as viable inflationary models\footnote{For early steps towards inflationary 6D models see \cite{HML}. } \cite{SIreviews}. For the system studied here this is accomplished using a simple flux-stabilization mechanism, that fixes all bulk properties except the overall volume modulus.

Incorporating the back-reaction of the branes in these solutions is the main feature new to this paper. It is important because it allows the explicit determination of how the extra-dimensional geometry responds to the choices made for a matter field, which we assume to be localized on one of the source branes. It also provides a mechanism for lifting the one remaining flat direction, through a codimension-two generalization of the Goldberger-Wise mechanism \cite{GW} of codimension-one Randall-Sundrum models.

In order to compute the back-reaction we extend to time-dependent geometries the bulk-brane matching conditions that were previously derived for codimension-two branes only in the limit of maximally symmetric on-brane geometries \cite{Cod2Matching, BBvN, BulkAxions, susybranes}. We then apply these conditions to the time-dependent bulk geometries to see how their integration constants are related to physical choices made for the dynamics of an `inflaton' scalar field that we assume to be localized on one of the source branes.

For the solutions we describe, the scale factor of the on-brane dimensions expands like $a(t) \propto t^p$, and our main interest is on the accelerating solutions (for which $p > 1$). The parameter $p$ is an integration constant for the bulk solution, whose value becomes related to the shape of the potential for the on-brane scalar. de Sitter solutions \cite{6DdS} are obtained in the limit $p \to \infty$, which corresponds to the limit where the on-brane scalar potential becomes independent of the inflaton.

What is most interesting is what the other dimensions do while the on-brane geometry inflates: their radius expands with a universal expansion rate, $r(t) \propto \sqrt t$, that is $p$-independent for any finite $p$. (By contrast, the extra dimensions do not expand at all for the special case of the de Sitter solutions.) The different expansion rates therefore cause the accelerated expansion of the on-brane directions to be faster than the growth of the size of the extra-dimensional directions; possibly providing the seeds of an understanding of why the on-brane dimensions are so much larger at the present epoch, in our much later universe.

Because the extra dimensions expand (rather than contract), the Kaluza-Klein mass scale falls with time, putting the solution deeper into the domain of validity of the low-energy semiclassical regime. Equivalently, the higher-dimensional gravity scale falls (in 4D Planck units) during the inflationary epoch. This opens up the intriguing possibility of reconciling a very low gravity scale during the present epoch with a potentially much higher gravity scale when primordial fluctuations are generated during inflation.

In the limit where the motion is adiabatic, we verify how the time-dependence of the full theory is captured by the solutions of the appropriate effective low-energy 4D theory. The 4D description of the inflationary models turns out to resemble in some ways an extended inflation model \cite{ExtInf}, though with an in-principle calculable potential for the Brans-Dicke scalar replacing the cosmological-constant sector that is usually assumed in these models.

The rest of this paper is organized as follows. The next section, \S2, summarizes the field equations and solutions that describe the bulk physics in the model of interest. A particular focus in this section is the time-dependence and the asymptotics of the solutions in the vicinity of the two source branes. These are followed in \S3\ by a description of the dynamics to be assumed of the branes, as well as the boundary conditions that are dictated for the bulk fields by this assumption. The resulting matching conditions are then used to relate the parameters of the bulk solution to the various brane couplings and initial conditions assumed for the brane-localized scalar field. \S4\ then describes the same solutions from the point of view of a 4D observer, using the low-energy 4D effective theory that captures the long-wavelength physics. The low-energy field equations are solved and shown to share the same kinds of solutions as do the higher-dimensional field equations, showing how the two theories can capture the same physics. Some conclusions and outstanding issues are discussed in \S5. Four appendices provide the details of the brane properties; the derivation of the time-dependent codimension-two matching conditions; and the dimensional reduction to the 4D effective theory.

\section{The bulk: action and solutions}

In this section we summarize the higher-dimensional field equations and a broad class of time-dependent solutions, whose properties are matched to those of the source branes in the next section. For definiteness we use the equations of 6D chiral gauged super-gravity \cite{NS} with flux-stabilized extra dimensions. The minimal number of fields to follow are the 6D metric, $g_{\ssM \ssN}$, and dilaton, $\phi$, plus a flux-stabilizing Maxwell potential, $A_\ssM$. Although other fields are present in the full theory, only these three need be present in the simplest flux-stabilized solutions \cite{SSs, ConTrunc}.

The action for these fields is
\be \label{BulkAction}
 S_\mathrm{bulk} = - \int \exd^6 x \sqrt{-g} \; \left\{ \frac1{2\kappa^2} \, g^{\ssM\ssN}
 \Bigl( \cR_{\ssM \ssN} + \pd_\ssM \phi \, \pd_\ssN \phi \Bigr)
 + \frac14 \, e^{-\phi} \cF_{\ssM\ssN} \cF^{\ssM\ssN}
 + \frac{2 \, g_\ssR^2}{\kappa^4} \, e^\phi \right\} \,,
\ee
where $\kappa^2 = 8\pi G_6 = 1/M_6^4$ defines the 6D Planck scale and $\cF = \exd \cA$ is the field strength for the Maxwell field, whose coupling is denoted by $g$. The coupling $g$ can be, but need not be, the same as the coupling $g_\ssR$ that appears in the scalar potential, since supersymmetry requires $g_\ssR$ must be the gauge coupling for a specific $U(1)_\ssR$ symmetry that does not commute with supersymmetry. $g$ would equal $g_\ssR$ if $\cA_\ssM$ gauges this particular symmetry, but need not otherwise.

The field equations coming from this action consist of the Einstein equation
\be \label{BulkEinsteinEq}
 \cR_{\ssM\ssN} + \partial_\ssM \phi \, \partial_\ssN \phi
  + \kappa^2  e^{-\phi} \cF_{\ssM \ssP} {\cF_\ssN}^\ssP
  - \left( \frac{\kappa^2}{8} \,
  e^{-\phi} \cF_{\ssP\ssQ} \cF^{\ssP \ssQ}
  - \frac{g_\ssR^2}{\kappa^2} \, e^\phi  \right)
  g_{\ssM\ssN} = 0 \,,
\ee
the Maxwell equation
\be \label{BulkMaxwellEq}
 \nabla_\ssM (e^{-\phi} \cF^{\ssM \ssN}) = 0 \,,
\ee
and the dilaton equation
\be \label{BulkDilatonEq}
 \Box \phi - \frac{2 \, g_\ssR^2 }{\kappa^2} \, e^\phi  + \frac{\kappa^2}4 \, e^{-\phi} \cF_{\ssM\ssN} \cF^{\ssM\ssN} = 0 \,.
\ee
Notice these equations are invariant under the rigid rescaling,
\be \label{scaleinv}
 g_{\ssM \ssN} \to \zeta \, g_{\ssM \ssN}
  \quad \hbox{and} \quad
  e^\phi \to \zeta^{-1} \, e^\phi \,,
\ee
with $\cA_\ssM$ unchanged, which ensures the existence of a zero-mode that is massless at the classical level, and much lighter than the generic KK scale once quantum effects are included.

\subsection{Bulk solutions}

The exact solutions to these equations we use for cosmology are described in \cite{scaling solutions} (see also \cite{CopelandSeto}). Their construction exploits the scale invariance of the field equations to recognize that exact time-dependent solutions can be constructed by scaling out appropriate powers of time from each component function.

\subsubsection*{Time-dependent ansatz}

Following \cite{scaling solutions} we adopt the following ansatz for the metric,
\be
 \label{metric-ansatz}
 \exd s^2 = (H_0\tau)^c \left\{ \left[ -e^{2\omega(\eta)} \exd\tau^2
 +e^{2 \alpha (\eta)} \delta_{ij} \exd x^i \exd x^j \right]
 + \tau^{2} \left[e^{2v(\eta)} \exd\eta^2 +
  e^{2 \beta(\eta)}\exd\theta^2\right] \right\} \,,
\ee
while the dilaton and Maxwell field are assumed to be
\be \label{dilaton-ansatz}
 e^\phi = \frac{e^{\varphi(\eta)} }{(H_0\tau)^{2+c}} \quad \hbox{and} \quad
 \cA_\theta = \frac{ A_\theta(\eta) }{H_0} \,.
\ee

The power of time, $\tau$, appearing in each of these functions is chosen to ensure that all of the $\tau$-dependence appears as a common factor in each of the field equations. The 6D field equations then reduce to a collection of $\tau$-independent conditions that govern the profiles of the functions $\varphi$, $\omega$, $\alpha$, $\beta$, $v$ and $A_\theta$. For later convenience we briefly digress to describe the properties of these profiles in more detail.

\subsubsection*{Radial profiles}

Explicitly, with the above ansatz the Maxwell equation becomes
\be
 A_\theta'' + \(\omega + 3 \alpha - \beta - v - \varphi \)' A_\theta' = 0 \,,
\ee
where primes denote differentiation with respect to the coordinate $\eta$. The dilaton equation similarly is
\be
 \varphi'' + \( \omega + 3 \alpha - v + \beta \)' \varphi'
 + (2+c)(1+2c) \, e^{2(v-\omega)}
 + \frac{\kappa^2}2 \, e^{-(2\beta + \varphi)}(A_\theta')^2
 -\frac{2g_\ssR^2}{\kappa^2 H_0^2} \, e^{2v+\varphi} = 0 \,.
\ee
The $\tau$-$\eta$ Einstein equation is first order in derivatives,
\be \label{eq:Einst}
 (2c+1) \, \omega' +3\alpha' + (2+c) \, \varphi' = 0 \,,
\ee
while the rest are second order
\ba
 \omega''+\(\omega+3\alpha -v +\beta \)' \omega'
 +\frac{\kappa^2} 4 \, e^{-(2 \beta + \varphi)}
 (A_\theta')^2 + \frac{g_\ssR^2}{\kappa^2 H_0^2} \, e^{2v+\varphi}
 -\left(c^2+\frac{5c}{2} +4 \right) \, e^{2(v-\omega)} &=&0\nn\\
 \beta'' + \(\omega+3\alpha -v +\beta\)' \beta'
 +\frac{3\kappa^2}4 \, e^{-(2\beta +\varphi)}(A_\theta')^2
 + \frac{g_\ssR^2}{\kappa^2 H_0^2} \, e^{2v+\varphi}
 -\frac12(c+2)(2c+1) \, e^{2(v-\omega)} &=&0 \nn\\
 \alpha'' + \(\omega+3\alpha -v+\beta\)' \alpha' - \frac{\kappa^2}4
 \, e^{-(2\beta +\varphi)}(A_\theta')^2
 +\frac{g_\ssR^2}{\kappa^2 H_0^2} \, e^{2v+\varphi}
 -\frac{c}{2}\,(2c+1) \, e^{2(v-\omega)} &=&0\nn\\
 \omega'' + 3\alpha'' + \beta'' + (\omega')^2
 +3(\alpha')^2 + (\beta')^2 + (\varphi')^2 -\(\omega+3\alpha +\beta\)'
 v'\qquad\qquad\qquad\qquad\quad&&\nn\\
 +\frac{3\kappa^2}4 \, e^{-(2\beta +\varphi)}(A_\theta')^2
 +\frac{g_\ssR^2}{\kappa^2 H_0^2} \, e^{2v+\varphi}
 -\frac12(c+2)(2c+1) \, e^{2(v-\omega)} &=&0 \,. \nn\\
\ea
One linear combination of these --- the `Hamiltonian' constraint for evolution in the $\eta$ direction --- also doesn't involve any second derivatives, and is given by
\ba \label{eq:Hamconst}
  &&(\varphi')^2 - 6\(\omega + \alpha + \beta\)' \alpha'
 - 2 \omega' \beta' \nn\\
 && \qquad\qquad +\frac{\kappa^2}2 \, e^{-(2\beta +\varphi)}(A_\theta')^2
 - \frac{4g_\ssR^2}{\kappa^2 H_0^2} \, e^{2v+\varphi}
 +4 (c^2+c+1) \, e^{2(v-\omega)}  = 0 \,.
\ea

As shown in \cite{scaling solutions}, these equations greatly simplify if we trade the four functions $\alpha$, $\beta$, $\omega$ and $\varphi$ for three new functions $\cx$, $\cy$ and $\cz$, using the redefinitions
\ba \label{XYZdef}
 \omega&=&-\frac\cx4+\frac\cy4+ \left( \frac{2+c}{2c} \right) \cz
 \,, \qquad
 \alpha = -\frac\cx4+\frac\cy4- \left( \frac{2+c}{6c} \right) \cz \,,\nn\\
 && \qquad \beta = \frac{3\cx}4+\frac\cy4+\frac\cz2 \quad
 \hbox{and} \quad
 \varphi = \frac\cx2-\frac\cy2-\cz \,.
\ea
Only three functions are needed to replace the initial four because these definitions are chosen to identically satisfy eq.~\pref{eq:Einst} which, for the purposes of integrating the equations in the $\eta$ direction, can be regarded as a constraint (because it doesn't involve any second derivatives). The function $v$ can be chosen arbitrarily by redefining $\eta$, and the choice
\be
  v = -\frac\cx4+\frac{5\cy}4+\frac\cz2 \,,
\ee
proves to be particularly simple \cite{scaling solutions}.

In terms of these variables the Maxwell equation becomes
\be
 A_\theta'' - 2\cx' A_\theta' = 0 \,,
\ee
the dilaton equation is
\be
 \(\frac12\cx-\frac12\cy-\cz\)''+(c+2)(2c+1) \, e^{2(\cy-\cz/c)} +
 \frac{\kappa^2}2 \, e^{-2\cx}(A_\theta')^2 - \frac{2g_\ssR^2}{\kappa^2 H_0^2} \, e^{2\cy}=0 \,,
\ee
and the remaining Einstein equations are
\ba
 \(-\frac14\cx+\frac14\cy+\frac{2+c}{2c}\cz\)''+\frac{\kappa^2} 4 \,
 e^{-2 \cx} (A_\theta')^2 + \frac{g_\ssR^2}{\kappa^2 H_0^2} \, e^{2\cy}
 -\left( c^2+\frac{5c}{2}+4 \right) \, e^{2(\cy-\cz/c)} &=&0\nn\\
 \( \frac34\cx+\frac14\cy+\frac12\cz \)'' +\frac{3\kappa^2}4 \,
 e^{-2\cx}(A_\theta')^2 + \frac{g_\ssR^2}{\kappa^2 H_0^2} \, e^{2\cy}
 -\frac12(c+2)(2c+1) \, e^{2(\cy-\cz/c)} &=&0\nn\\
 \(\frac14\cx+\frac14\cy-\frac{2+c}{6c}\cz \)'' - \frac{\kappa^2}4
 \, e^{-2\cx}(A_\theta')^2 +\frac{g_\ssR^2}{\kappa^2 H_0^2} \, e^{2\cy}
 -\frac12c(2c+1) \, e^{2(\cy-\cz/c)} &=&0 \nn\\
 \(-\frac14\cx+\frac54\cy+\frac12\cz\)''
 +(\cx')^2-(\cy')^2+\frac43 \, \frac{1+c+c^2}{c^2} \, (\cz')^2
 \qquad\qquad\qquad\qquad&&\nn\\
 +\frac{3\kappa^2}4 \, e^{-2\cx}(A_\theta')^2
 +\frac{g_\ssR^2}{\kappa^2 H_0^2} \, e^{2\cy}
 -\frac12(c+2)(2c+1) \, e^{2(\cy-\cz/c)} &=&0 \,. \nn\\
\ea

The combination of twice the second Einstein equation plus the Dilaton equation is completely independent of $\cy$ and $\cz$. This combination and the Maxwell equation can be exactly integrated, giving
\ba \label{eq:chisoln}
 A_\theta &=& q \int\exd\eta \; e^{2\cx}\nn\\
 e^{-\cx} &=& \left( \frac{\kappa \, q}{\lambda_1} \right)
 \cosh\left[ \lambda_1(\eta-\eta_1) \right],
\ea
where $q$, $\lambda_1$ and $\eta_1$ are integration constants.

The remaining field equations then reduce to
\ba
 \label{bulkXY}
 \cy''+\frac{4g_\ssR^2}{\kappa^2 H_0^2} \, e^{2\cy} - 4 (1+c+c^2)
 \, e^{2\cy-2\cz/c}&=&0\nn\\
 \hbox{and} \qquad
 \cz'' - 3c\, e^{2\cy-2\cz/c}&=&0 \,,
\ea
together with the first-order constraint, eq.~\pref{eq:Hamconst}, that ensures that only two of the `initial conditions' --- $\cx'$, $\cy'$ and $\cz'$ --- are independent.

\subsubsection*{Asymptotic forms}

With these coordinates the singularities of the metric lie at $\eta \to \pm \infty$, which is interpreted as the position of two source branes. We now pause to identify the asymptotic forms to be required by the metric functions as these branes are approached.

There are two physical conditions that guide this choice. First, we wish the limits $\eta \to \pm \infty$ to represent codimension-two points, rather than codimension-one surfaces, and so require $e^{2\beta} \to 0$ in this limit. In addition, we require the two extra dimensions to have finite volume, which requires $e^{\beta + v} \to 0$.

In Appendix \ref{app:AsForm} we argue, following \cite{scaling solutions}, that these conditions require both $\cy''$ and $\cz''$ must vanish in the limit $\eta \to \pm \infty$, and so
the functions $\cy$ and $\cz$ asymptote to linear functions of $\eta$ for large $|\eta|$:
\be \label{eq:YZbcs}
 \cy \to \cy_\infty^\pm \mp\lambda_2^\pm\eta \quad \hbox{and} \quad
 \cz \to \cz_\infty^\pm \mp\lambda_3^\pm\eta \quad \hbox{as} \quad
 \eta \to \pm \infty \,,
\ee
where $\cy_\infty^\pm$, $\cz_\infty^\pm$, $\lambda_2^\pm$ and $\lambda_3^\pm$ are integration constants. The signs in eqs.~\pref{eq:YZbcs} are chosen so that $\lambda_2^\pm$ and $\lambda_3^\pm$ give the outward-pointing normal derivatives: {\em e.g.} $\lim_{\eta \to \pm \infty} N \cdot \partial \cy = \lambda_2^\pm$, where $N_\ssM$ denotes the outward-pointing unit normal to a surface at fixed $\eta$.

Not all of the integration constants identified to this point are independent of one another, however. In particular, the asymptotic form as $\eta \to +\infty$ can be computed from that at $\eta \to - \infty$ by integrating the field equations, and so cannot be independently chosen. In principle, given a value for $c$ and for all of the constants $\lambda_i^+$, $\cx_\infty^+$, $\cy_\infty^+$ and $\cz_\infty^+$, integration of the bulk field equations yields the values for $\lambda_i^-$, $\cx_\infty^-$, $\cy_\infty^-$ and $\cz_\infty^-$.

In addition, the integration constants need not all be independent even restricting our attention purely to the vicinity of only one of the branes. There are several reasons for this. One combination of these field equations --- the `Hamiltonian' constraint, eq.~\pref{eq:Hamconst} --- imposes a condition\footnote{If this constraint is satisfied as $\eta \to -\infty$, the equations of motion automatically guarantee it also holds as $\eta \to + \infty$.} that restricts the choices that can be made at $\eta \to - \infty$,
\be
 \label{eq:powersconstraint}
 (\lambda_2^\pm)^2 = \lambda_1^2 + \frac43 \left( \frac{1+c+c^2}{c^2} \right) (\lambda_3^\pm)^2 \,.
\ee
Also, it turns out that the constants $\cx_\infty^\pm$ are not independent of the other parameters describing the bulk solution, like the flux-quantization integer $n$ to be discussed next.

Next, flux quantization for the Maxwell field in the extra dimensions also imposes a relation amongst the integration constants. In the absence of brane sources, flux quantization implies \cite{scaling solutions}
\be
 \frac n{g} = \frac{q}{H_0} \int_{-\infty}^\infty\exd\eta \; e^{2\cx}
 = \frac{\lambda_1^2}{q\kappa^2 H_0} \, \int_{-\infty}^\infty\exd\eta \, \cosh^{-2}\left[\lambda_1(\eta-\eta_1)\right]
 = \frac{2\lambda_1}{q\kappa^2 H_0} \,,
\ee
where $n$ is an integer. This gets slightly modified when branes are present, if the branes are capable of carrying a brane-localized Maxwell flux \cite{BulkAxions, susybranes} (as is the case in particular for the branes considered in \S3, below). In this case the flux-quantization condition is modified to
\be
 \label{eq:bulkfluxquant}
 \frac n{g} = \sum_b \frac{\Phi_b(\phi)}{2\pi}
 +\frac{2\lambda_1}{q\kappa^2 H_0} \,,
\ee
where $\Phi_b$ is the flux localized on each brane. (More on this when we discuss brane properties in more detail in \S3.)

Finally, since the above solutions transform into one other under constant shifts of $\eta$, we may use this freedom to reparameterize $\eta \to \eta+\eta_1$ to eliminate $\eta_1$, in which case
\be
 e^{-\cx}=\frac{\kappa\, q}{\lambda_1} \, \cosh(\lambda_1\eta)
 = \frac{4\pi g}{\kappa H_0(2\pi n-g\sum_b\Phi_b)} \, \cosh(\lambda_1\eta).
\ee
{}From this we see that the asymptotic form for $\cx$ is
\be
 \cx\to\cx_\infty^\pm\mp\lambda_1\eta\,,
\ee
with
\be
 \cx_\infty^\pm = \ln\left[ \kappa H_0\(
 \frac n{g}-\sum_b\frac{\Phi_b}{2\pi}\) \right] \,.
\ee
This shows explicitly how $\cx_\infty^\pm$ is related to other integration constants.

All told, this leaves $c$, $H_0$, $\lambda_2^-$, $\lambda_3^-$, $\cy^-$ and $\cz^-$ (or, equivalently, $c$, $H_0$, $\lambda_2^+$, $\lambda_3^+$, $\cy^+$ and $\cz^+$) as the six independent integration constants of the bulk solution. These we relate to brane properties in subsequent sections.

\subsection{Interpretation as 4D cosmology}

In order to make contact with the cosmology seen by a brane-localized observer, we must put the 4D metric into standard Friedmann-Lemaitre-Robertson-Walker (FLRW) form. In particular, we should do so for the 4D Einstein-frame metric, for which the 4D Planck scale is time-independent.

\subsubsection*{4D Einstein frame}

Recall the 6D metric has the form
\ba
 g_{\ssM \ssN} \, \exd x^\ssM\exd x^\ssN &=& (H_0\tau)^c \Bigl\{ \left[
 -e^{2\omega} \exd\tau^2
 + e^{2 \alpha} \delta_{ij} \,\exd x^i \exd x^j \right]
 + \tau^{2} \left[ e^{2v} \exd\eta^2
 + e^{2\beta} \exd\theta^2 \right]  \Bigr\} \nn\\
 &=& \hat g_{\mu\nu}\exd x^\mu \exd x^\nu + \frac{(H_0\tau)^{2+c}}{H_0^2}
 \left[e^{2v}\exd\eta^2
 +e^{2\beta} \exd\theta^2 \right] \,,
\ea
and denote by $\hat R_{\mu \nu}$ the Ricci tensor constructed using $\hat g_{\mu\nu}$. In terms of these, the time dependence of the 4D Einstein-Hilbert term is given by
\be
 \frac{1}{2 \kappa^2} \sqrt{-g} \; g^{\ssM \ssN} R_{\ssM \ssN}
 = \frac{1}{2 \kappa^2 H_0^2} \sqrt{-\hat g} \; \hat g^{\mu\nu} \hat R_{\mu\nu} \;
 e^{\beta +v}(H_0 \tau)^{2+c} + \cdots \,.
\ee
This time dependence can be removed by defining a new 4D Einstein-frame metric
\be
 \tilde g_{\mu\nu} = (H_0\tau)^{2+c} \hat g_{\mu\nu} \,,
\ee
whose components are
\be
 \tilde g_{\mu\nu} \, \exd x^\mu \exd x^\nu = (H_0\tau)^{2+2c}
 \left[ -e^{2\omega} \exd \tau^2 + e^{2\alpha} \delta_{ij} \,
 \exd x^i\exd x^j\right] \,.
\ee

\subsubsection*{FLRW time}

FLRW time is defined for this metric by solving $\exd t = \pm (H_0 \tau)^{1 + c} \exd \tau$. There are two cases to consider, depending on whether or not $c=-2$. If $c \ne -2$, then
\be
 H_0 t =  \frac{|H_0 \tau|^{2+c}}{|2+c|}
 \qquad ( \hbox{if} \quad c \ne -2)\,,
\ee
where the sign is chosen by demanding that $t$ increases as $\tau$ does. (If $c < -2$ then $t$ rises from 0 to $\infty$ as $\tau$ climbs from $-\infty$ to 0.) This puts the 4D metric into an FLRW-like form
\be \label{FLRWwarpedform}
 \tilde g_{\mu\nu} \, \exd x^\mu \exd x^\nu = - e^{2\omega} \, \exd t^2
 + a^2(t) \,  e^{2\alpha} \delta_{ij} \, \exd x^i\exd x^j \,,
\ee
where
\be
 a(t) =  ( |c+2| \, H_0 t)^p \quad \hbox{with} \quad
 p = \frac{1+ c}{2+c}
 \qquad ( \hbox{if} \quad c \ne -2)\,.
\ee

\FIGURE[ht]{
\epsfig{file=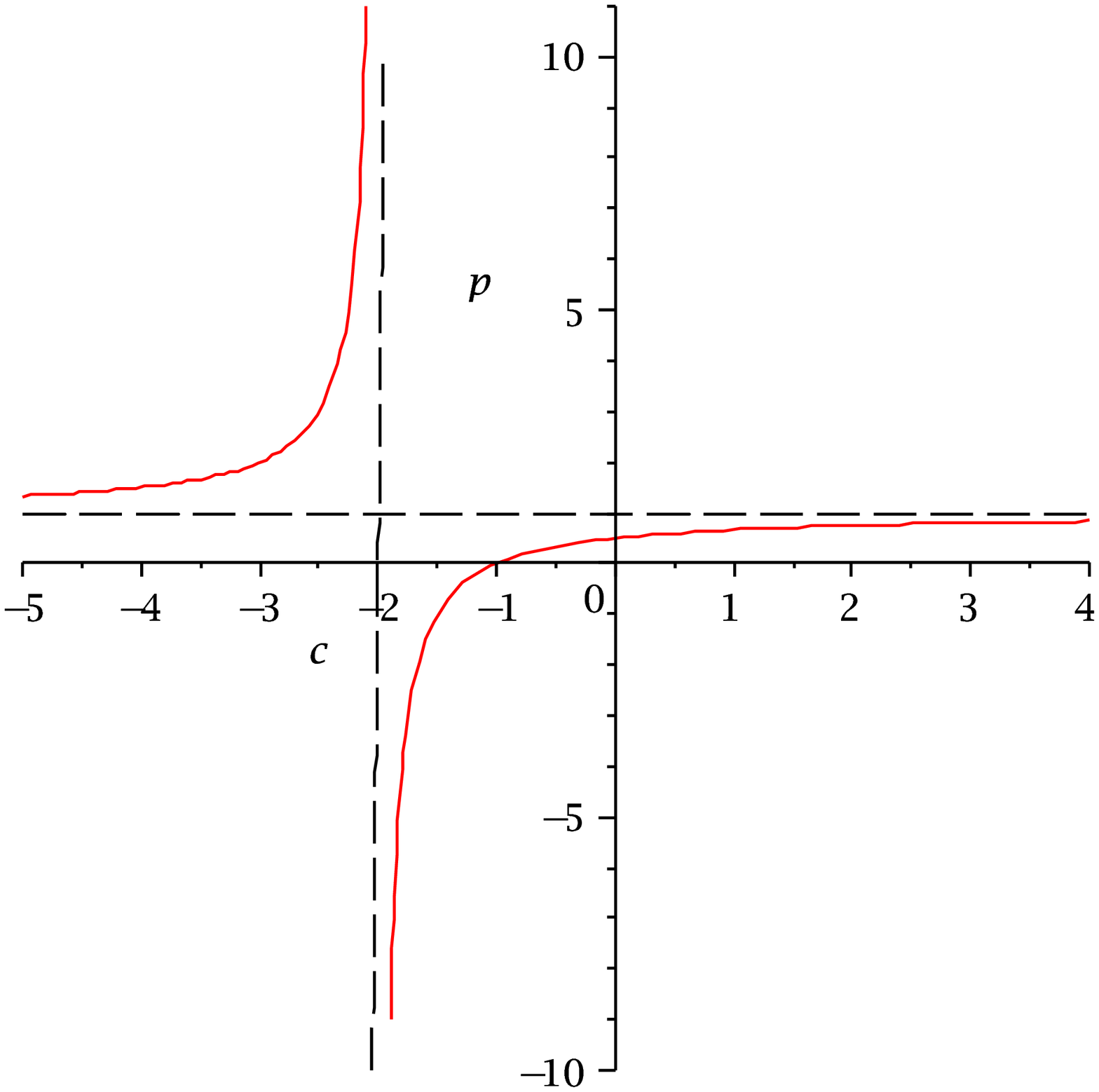,angle=0,width=0.35\hsize}
\caption{A plot of the power, $p$, controlling the scale factor's expansion, vs the parameter $c$ appearing in the higher-dimensional ansatz.
} \label{fig:pvsc} }

Notice that $p > 1$ if $c < -2$, with $p \to 1$ as $c \to - \infty$ and $p \to + \infty$ when $c \to -2$ from below (see fig.~\ref{fig:pvsc}). This describes an accelerated power-law expansion, resembling the power-law expansion of `extended inflation' \cite{ExtInf} for which $\ddot a/a = p\,(p-1)/t^2 > 0$. Similarly, $p < 0$ if $-2 < c < -1$, with $p \to 0$ as $c \to - 1$ and $p \to - \infty$ as $c \to -2$ from above. Since $p < 0$ this describes a 4D universe that contracts as $t$ increases. Finally $0 < p < 1$ if $c > -1$, climbing monotonically from zero with increasing $c$ until $p \to 1$ as $c \to + \infty$. Since $\ddot a/a < 0$, this describes decelerated expansion.

If $c=-2$, we instead define
\be
 H_0 t = - \ln|H_0\tau|
 \qquad ( \hbox{if} \quad c = -2) \,,
\ee
in which case the FLRW metric again takes the form of eq.~\pref{FLRWwarpedform}, with
\be
 a(t) = e^{H_0 t}
 \qquad ( \hbox{if} \quad c = -2)\,.
\ee
This is the limiting case of the de Sitter-like solutions, found in \cite{6DdS}.

It may seem a surprise to find de Sitter solutions, given the many no-go results \cite{dSnogo}, however these de Sitter solutions thread a loop-hole in the no-go theorems. The loop-hole is the benign-looking assumption of compactness: that integrals of the form $I := \int \exd^n x \; \sqrt{g} \; \Box X$ must vanish, where $X$ is a suitable combination of bulk fields. This assumption is violated due to the back-reaction of the branes, since this can force the bulk fields to become sufficiently singular near the branes to contribute nonzero contributions to integrals like $I$ \cite{6DdS, BBvN}.

\subsubsection*{$t$-dependence of other bulk fields}

Recalling that the extra-dimensional metric has the form
\be
 \exd s^2 = \frac{|H_0 \tau|^{2+c}}{H_0^2} \left( e^{2v } \exd \eta^2 + e^{2\beta} \exd \theta^2 \right) \,,
\ee
we see that the linear size of the extra dimensions is time-independent if $c = -2$, but otherwise behaves as
\be
 r(t) \propto  \frac{|H_0 \tau|^{1 + c/2}}{H_0}
 = \frac{(|c+2| \, H_0 t)^{1/2}}{H_0}
 \qquad ( \hbox{if} \quad c \ne -2) \,.
\ee
This shows that the extra dimensions universally grow as $r \propto \sqrt t$ for any $c \ne -2$. In particular $r(t)$ grows even if $a(t) \propto t^p$ shrinks (which happens when $p < 0$: {\em i.e.} when $-2 < c < -1$). When $a(t)$ grows, it grows faster than $r(t)$ whenever $p > \frac12$, and which is true for both $c < -2$ and for $c > 0$. It is true in particular whenever the expansion of the on-brane directions accelerates ({\em i.e.} when $p > 1$). When $0 < p < \frac12$ (and so $-1 < c < 0$) it is the extra dimensions that grow faster.

Another useful comparison for later purposes is between the size of $r(t)$ and the 4D Hubble length, $H^{-1}(t)$. Since neither $r$ nor $H$ depends on time when $c = -2$, this ratio is also time-independent in this limit. But for all other values of $c$, the Hubble scale is given by $H := \dot a/a = p/t$, with $p = (c+1)/(c+2)$, as above. Consequently, the ratio of $H$ to the KK scale, $m_\KK = 1/r$, is given by
\be
 H(t)\, r(t) \propto  \frac{|c+1|}{( |c+2| \, H_0 t)^{1/2}} \,,
\ee
and so decreases as $t$ evolves.

The dilaton also has a simple time-dependence when expressed as a function of $t$. It is time-dependent if $c = -2$, but otherwise evolves as
\be
 e^\phi \propto \frac{1}{(H_0 \tau)^{2+c}} \propto \frac{1}{t}
 \qquad ( \hbox{if} \quad c \ne -2) \,,
\ee
which shows that $r^2 e^\phi$ remains independent of $t$ for all $c$. Notice that this implies that evolution takes us deeper into the regime of weak coupling, since it is $e^\phi$ that is the loop-counting parameter of the bulk supergravity \cite{susybranes, TNCC}.

\section{Brane actions and bulk boundary conditions}

It is not just the geometry of the universe that is of interest in cosmology, but also how this geometry responds to the universal energy distribution. So in order to properly exploit the above solutions to the field equations it is necessary to relate its integration constants to the physical properties of the matter that sources it. In the present instance this requires specifying an action for the two source branes that reside at $\eta \to \pm \infty$.

To this end we imagine one brane to be a spectator, in the sense that it does not involve any on-brane degrees of freedom. Its action therefore involves only the bulk fields, which to lowest order in a derivative expansion is\footnote{Although nominally involving one higher derivative than the tension term, the magnetic coupling, $\Phi$, describes the amount of flux that can be localized on the brane \cite{BulkAxions, susybranes}, and can be important when computing the energetics of flux-stabilized compactifications in supergravity because of the tendency of the tension to drop out of this quantity \cite{susybranes, TNCC}. We follow here the conventions for $\Phi$ adopted in \cite{BulkAxions, susybranes}, which differ by a factor of $e^{-\phi}$ from those of \cite{TNCC}.}
\be \label{eq:Sspec}
 S_s = - \int \exd^4x \sqrt{-\gamma} \; \left\{ T_s
   - \frac12 \, \Phi_s \, e^{-\phi} \, \epsilon^{mn} \cF_{mn} + \cdots \right\} \,.
\ee
Here $T_s$ and $\Phi_s$ are dimensionful parameters, $\gamma_{mn}$ is the induced on-brane metric, and $\epsilon^{mn}$ is the antisymmetric tensor defined on the two dimensions transverse to the brane. Physically, $T_s$ denotes the tension of the spectator brane, while the magnetic coupling, $\Phi_s$, has the physical interpretation of the amount of flux that is localized at the brane \cite{BulkAxions, susybranes} (see Appendix \ref{app:FluxQ}).

To provide the dynamics that drives the bulk time dependence we imagine localizing a scalar field --- or inflaton, $\chi$ --- on the second, `inflaton', brane with action
\be \label{eq:Sinf}
 S_i = - \int \exd^4x \sqrt{- \gamma}  \; \left\{ T_i +
  f(\phi) \Bigl[ \gamma^{\mu\nu} \pd_\mu \chi \pd_\nu \chi
  + V(\chi) \Bigr] - \frac12 \, \Phi_i \, e^{-\phi}
  \, \epsilon^{mn} \cF_{mn} + \cdots \right\} \,.
\ee
As before $T_i$ and $\Phi_i$ denote this brane's tension and bulk flux, both of which we assume to be independent of the bulk dilaton, $\phi$. In what follows we assume the following explicit forms,
\be
 f(\phi) = e^{-\phi}
 \quad \hbox{and} \quad
 V(\chi) = V_0 +  V_1 \, e^{\zeta \chi}
 + V_2 \, e^{2\zeta \chi} + \cdots \,,
\ee
but our interest is in the regime where the term $V_1 \, e^{\zeta \chi}$ dominates all the others in $V(\chi)$, and so we choose the coefficients $V_k$ appropriately. These choices --- $f = e^{-\phi}$ and $V = V_1 \, e^{\zeta \, \chi}$, as well as the $\phi$-independence of $T_s$, $T_i$, $\Phi_s$ and $\Phi_i$ --- are special because they preserve the scale invariance, eq.~\pref{scaleinv}, of the bulk equations of motion.

As we see below, these choices for the functions $f(\phi)$ and $V(\chi)$ are required in order for the equations of motion for $\chi$ to be consistent with the power-law time-dependence we assume above for the solution in the bulk. In order to see why this is true, we require the matching conditions that govern how this action back-reacts onto the properties of the bulk solution that interpolates between the two branes. This requires the generalization to time-dependent systems of the codimension-two matching conditions worked out elsewhere \cite{Cod2Matching, BBvN} for the special case of maximally symmetric on-brane geometries. These matching conditions generalize the familiar Israel junction conditions that relate bulk and brane properties for codimension-one branes, such as those encountered in Randall-Sundrum type models \cite{RS}.

\subsection{Time-dependent brane-bulk matching}

When the on-brane geometry is maximally symmetric --- {\em i.e.} flat, de Sitter or anti-de Sitter --- the matching conditions for codimension-two branes are derived in refs.~\cite{Cod2Matching} (see also \cite{PST}), and summarized with examples in ref.~\cite{BBvN}. In Appendix \ref{matchingderivation} we generalize these matching conditions to the case where the on-brane geometry is time-dependent, in order to apply it to the situation of interest here. In this section we describe the result of this generalization.

For simplicity we assume axial symmetry in the immediate vicinity of the codimension-2 brane, with $\theta$ being the coordinate labeling the symmetry direction and $\rho$ labeling a `radial' off-brane direction, with the brane located at $\rho = 0$. We do not demand that $\rho$ be proper distance, or even that $\rho$ be part of a system of orthogonal coordinates. However we do assume that there exist coordinates for which there are no off-diagonal metric components that mix $\theta$ with other coordinates: $g_{a \theta} = 0$. With those choices, the matching conditions for the metric are similar in form to those that apply in the maximally symmetric case:
\be \label{eq:cod2matching-g}
 - \frac12 \left[ \sqrt{g_{\theta\theta}}
 \, \left(K^{mn}-K  \proj^{mn}\right) - {\rm flat} \right]
 = \frac{\kappa^2}{2\pi} \, \frac1{\sqrt{-\gamma}}
 \, \frac{\delta S_b}{\delta g_{mn}} \,,
\ee
while those for the dilaton and Maxwell field are
\be \label{eq:cod2matching-phi}
 - \sqrt{g_{\theta\theta}}  \, N^m\nabla_m\phi
 = \frac{\kappa^2}{2\pi} \, \frac1{\sqrt{-\gamma}}
 \, \frac{\delta S_b}{\delta\phi} \,,
\ee
and
\be \label{eq:cod2matching-A}
 - \sqrt{g_{\theta\theta}}  \, e^{-\phi} \, N_mF^{mn}
 = \frac{\kappa^2}{2\pi} \, \frac1{\sqrt{-\gamma}} \,
 \frac{\delta S_b}{\delta A_n} \,.
\ee
Here the action appearing on the right-hand-side is the codimension-two action, such as eq.~\pref{eq:Sspec} or \pref{eq:Sinf}, and `flat' denotes the same result for a metric without a singularity at the brane position. We define the projection operator $\proj^m_n = \delta^m_n - N^m N_n$, where $N^m$ is the unit normal to the brane, pointing into the bulk. The induced metric $\gamma_{mn}$ is the projection operator restricted to the on-brane directions, and has determinant $\gamma$.

In principle the indices $m,n$ in eqs.~\pref{eq:cod2matching-g} run over all on-brane\footnote{When the metric has off-diagonal components mixing $\rho$ and brane directions, then $m,n$ also run over $\rho$. In our metric ansatz, those matching conditions vanish identically.} coordinates as well as $\theta$, and this might seem to present a problem since the codimension-2 action is not normally expressed as a function of $\theta$, since this is a degenerate coordinate at the brane position. However, the $\theta - \theta$ matching condition is never really required, because it is not independent of the others. Its content can instead be found from the others by using the `Hamiltonian' constraint, eq.~\pref{eq:Hamconst}, in the near-brane limit \cite{Cod2Matching, BBvN, otheruvcaps}.

\subsubsection*{Specialization to the bulk solutions}

Specialized to the geometry of our bulk ansatz, the above considerations lead to the following independent matching conditions for the inflationary brane. Writing the 4D on-brane coordinates as $\{ x^\mu \} = \{ t, x^i \}$, the $tt$, $ij$ and dilaton matching conditions become
\ba \label{eq:matchingform}
 \Bigl[ e^{\beta-v}(\partial_n \beta + 3 \partial_n \alpha)  \Bigr]_b
 &=& 1- \frac{\kappa^2}{2\pi} \left\{ T - H_0\Phi e^{-\varphi-v-\beta}A_\theta' + f(\phi) \left[
 -\pd_\tau \chi \, \pd^\tau \chi + V(\chi) \right] \right\} \nn\\
 \Bigl[ e^{\beta-v}(\partial_n \beta + 2 \partial_n \alpha
 + \partial_n \omega)\Bigr] &=&1 -\frac{\kappa^2}{2\pi} \, \left\{ T
 - H_0\Phi e^{-\varphi-v-\beta}A_\theta' + f(\phi) \left[ \pd_\tau \chi \, \pd^\tau \chi
 + V(\chi) \right] \right\} \nn\\
 \Bigl[ e^{\beta-v} \partial_n \phi \Bigr] &=& \frac{\kappa^2}{2\pi} \, \left(  f'(\phi) \left[ \pd_\tau\chi \, \pd^\tau\chi + V(\chi) \right] + H_0\Phi e^{-\varphi-v-\beta}A_\theta' \right) \,,
\ea
with $\partial_n = \pm \partial_\eta$ denoting the inward-pointing (away from the brane) radial derivative, and both sides are to be evaluated at the brane position --- {\em i.e.} with bulk fields evaluated in the limit\footnote{As we see below, any divergences in the bulk profiles in this near-brane limit are to be absorbed in these equations into renormalizations of the parameters appearing in the brane action.} $\eta \to \mp \infty$. In these equations $f'$ denotes $\exd f/\exd \phi$ while $A'_\theta = \partial_\eta A_\theta = F_{\eta \theta}$.

\subsubsection*{Consistency with assumed time-dependence}

We first record what $f(\phi)$ and $V(\chi)$ must satisfy in order for the matching conditions, eqs.~\pref{eq:matchingform}, to be consistent with the time-dependence assumed for the bulk cosmological solutions of interest here. Evaluating the left-hand side of the matching conditions, eqs.~\pref{eq:matchingform} using the ans\"atze of eqs.~\pref{metric-ansatz} and \pref{dilaton-ansatz} shows that they are time-independent. The same must therefore also be true of the right-hand side.

We choose $f(\phi)$ and $V(\chi)$ by demanding that the time-dependence arising due to the appearance of $\phi$ on the right-hand side cancel with time-dependence of the $\chi$-dependent pieces. Comparing the bottom two equations of \pref{eq:matchingform} then shows that the time-dependence of $f(\phi)$ and $f'(\phi)$ must be the same, and so $f(\phi) = C e^{k\phi}$ for some constants $C$ and $k$. The scale $C$ can be absorbed into the normalization of $\chi$, and so is dropped from here on.

Similarly, comparing the top two of eqs.~\pref{eq:matchingform} shows that the quantity $g^{\tau \tau} \partial_\tau \chi \, \partial_\tau \chi$ must scale with time in the same way as does $V(\chi)$. Furthermore, any scaling of $\chi$ with time must satisfy the $\chi$ equation of motion, found by varying the brane action with respect to $\chi$:
\be \label{eq:chieqn}
 \pd_\mu \left[ \sqrt{- \gamma} \; e^{k\phi} \pd^\mu \chi \right] - \sqrt{- \gamma} \; e^{k\phi} V'(\chi) = 0 \,.
\ee
Specialized to a homogeneous roll, $\chi=\chi(\tau)$, this simplifies to
\be \label{eq:chieqntau}
 \pd_\tau \left[ (H_0 \tau)^{2c} (H_0 \tau)^{-k(c+2)} e^{-2\omega}(H_0\tau)^{-c} \pd_\tau \chi \right] + (H_0 \tau)^{2c} (H_0 \tau)^{-k(c+2)} V'(\chi) = 0 \,.
\ee

All of these conditions are satisfied provided we assume a potential of the form
\be
 V(\chi) = V_1 \, e^{\zeta \chi} \,,
\ee
and an inflaton solution of the form
\be \label{chisoln}
 \chi = \chi_0+\chi_1\ln|H_0\tau| \,,
\ee
since in this case the time-dependence of the $\chi$ field equation factors. In what follows it is notationally useful to define $\hat V_1 := V_1 \, e^{\zeta \chi_0}$, allowing eqs.~\pref{eq:chieqn} and \pref{eq:chieqntau} to be rewritten as
\be
\label{inflaton-eom}
 H_0^2 e^{-2\omega} = \frac{\hat V_1 \zeta }{\chi_1(3+2\zeta \chi_1)}\,.
\ee
Notice that if $\zeta\chi_1 > 0$ then $V_1$ must also be non-negative.

In this case the conditions that $\pd_\tau \chi \pd^\tau \chi$ and $V(\chi)$ scale like $e^{-k\phi}$ boil down to
\be
 (H_0\tau)^{-c-2} \propto \tau^{\zeta \chi_1}  \propto \tau^{k(c+2)}  \,,
\ee
and so consistency between the scaling solutions and the matching condition implies $k = -1$, and so $f(\phi) = e^{-\phi}$ as anticipated earlier. It also determines the bulk time exponent $c$ in terms of brane properties:
\be \label{ceqn}
 \zeta \chi_1 = - (c+2) \,.
\ee

\subsection{Relation between brane parameters and physical bulk quantities}

We now use the above tools to establish more precisely the connection between brane properties and the physical characteristics of the bulk geometry.

\subsubsection*{Determination of integration constants}

Specializing the matching to the choices $f(\phi) = e^{-\phi}$ and $V(\chi) =  V_1 e^{\zeta \chi}$, and using the $\tau$-dependence of the bulk and brane fields described in \S2, gives the matching conditions in a form that determines the bulk integration constants in terms of properties of the two branes.

Consider first the spectator brane, for which the matching conditions are
\ba \label{redmatching}
 e^{\beta-v} \( \lambda_2^+ - \frac{\lambda_3^+}c \)
 &=& 1 - \frac{\kappa^2 T_s}{2\pi} + \frac{\kappa^2}{2\pi}H_0\Phi_se^{-\varphi-v-\beta}A_\theta'\nn\\
 e^{\beta-v} \( \lambda_2^+ + \frac{1+2c}{3c} \, \lambda_3^+ \)
 &=& 1 - \frac{\kappa^2T_s}{2\pi} + \frac{\kappa^2}{2\pi}H_0\Phi_se^{-\varphi-v-\beta}A_\theta' \\
 e^{\beta-v} \( \lambda_1 - \lambda_2^+ - 2 \lambda_3^+ \)
 &=& \frac{\kappa^2}{\pi}H_0\Phi_se^{-\varphi-v-\beta}A_\theta' \,, \nn
\ea
with all quantities evaluated at $\eta \to + \infty$. The difference between the first two of these implies
\be
 \lambda_3^+ = 0\,,
\ee
for the asymptotic geometry near the spectator brane, which also implies\footnote{ From the constraint alone,  $\lambda_1 = - \lambda_2^+$ is also allowed. The requirement of codimensions-2 branes together with finite volume excludes this possibility. For details, see appendix \ref{app:AsForm}} $\lambda_1=\lambda_2^+$ once the bulk constraint, eq.~\pref{eq:powersconstraint}, is used. This is then inconsistent with the third matching condition at this brane unless we also choose the spectator brane to contain no flux, $\Phi_s=0$. Given this, the matching conditions then degenerate into the usual defect-angle/tension relation \cite{TvsA}, which for the coordinates used here reads
\be
 \lambda_1=\lambda_2^+=e^{v-b}\(1-\frac{\kappa^2 T_s}{2\pi}\) \,.
\ee
This summarizes the near-brane geometry for a pure-tension brane for which $T_s$ does not depend on $\phi$.

Next consider the inflaton brane, for which matching implies
\ba
 e^{\beta-v} \( \lambda_2^- - \frac{\lambda_3^-}c \)
 &=& 1- \frac{\kappa^2}{2\pi} \, e^{-\varphi} \left[  e^{-2\omega}
 (H_0\chi_1)^2 + \hat V_1 - H_0 \, \Phi_i \, e^{-v-\beta}A_\theta'\right] - \frac{\kappa^2 T_i}{2\pi} \nn\\
e^{\beta-v} \( \lambda_2^- + \frac{1+2c}{3c} \, \lambda_3^- \)
&=& 1- \frac{\kappa^2}{2\pi} \, e^{-\varphi} \left[ - e^{-2\omega}
 (H_0\chi_1)^2 + \hat V_1 - H_0 \, \Phi_i \, e^{-v-\beta}A_\theta'\right] - \frac{\kappa^2 T_i}{2\pi} \nn\\
 e^{\beta-v} \( \lambda_1 - \lambda_2^- - 2 \lambda_3^- \)
 &=& \frac{\kappa^2}{\pi} \, e^{-\varphi} \left[
 e^{-2\omega}(H_0\chi_1)^2 - \hat V_1 + H_0 \, \Phi_i \, e^{-v-\beta} A_\theta'\right] \,,
\ea
with the fields evaluated at $\eta \to - \infty$. Using the first two matching conditions to eliminate $\lambda_2^-$, and using eqs.~\pref{inflaton-eom} and \pref{ceqn} to eliminate $H_0$ and $c$ allows the isolation of $\lambda_3^-$, giving
\be
 \label{lambda3matching}
 e^{\beta-v}\lambda_3^- = \frac{\kappa^2 \hat V_1}{2\pi}
 \( \frac{6+3 \, \zeta \chi_1}{3 + 2 \, \zeta \chi_1}\) \, e^{-\varphi}\,.
\ee

In general, matching for the inflaton brane is more subtle, since for it the above matching conditions typically diverge when evaluated at the brane positions. As usual \cite{Bren}, this divergence is absorbed into the parameters of the brane action, as we now briefly sketch.

\subsubsection*{Brane renormalization}

In general, in the near-brane limit $\beta-v=\cx-\cy$ varies linearly with $\eta$, approaching $\cx_\infty^\pm - \cy_\infty^\pm  \mp (\lambda_1 - \lambda_2^-) \, \eta$ as $\eta \to \pm \infty$. This shows that unless $\lambda_2^\pm=\lambda_1$ (which with the constraint, eq.~\pref{eq:powersconstraint}, then implies $\lambda_3^\pm=0$), the left-hand sides of the above matching conditions diverge. These divergences are generic to codimension-two and higher sources, as is familiar from the divergence of the Coulomb potential at the position of any source (codimension-3) point charges (in 3 space dimensions).

We absorb these divergences into renormalizations of the brane parameters, which in the present instance are $V_1$, $\zeta$, $T_i$ and $\Phi_i$, together with a wave-function renormalization of the on-brane field, $\chi$ (which for the present purposes amounts to a renormalization of $\chi_1$). To this end we regularize the matching conditions, by evaluating them at a small but nonzero distance away from the brane --- {\em i.e.} for $|\eta| = 1/\epsilon$ very large --- and assign an $\epsilon$-dependence to the couplings in such a way as to ensure that the renormalized results are finite as $\epsilon \to 0$. This is a meaningful procedure because the values of these parameters are ultimately determined by evaluating physical observables in terms of them, and measuring the values of these observables. Ultimately all of the uncertainties associated with the $\epsilon$ regularization cancel once the renormalized parameters are eliminated in this way in terms of observables, since a theory's predictive value is in the correlations it implies among the values of these observables.

In this section we (temporarily) denote the resulting renormalized (finite) brane parameters by a bar, {\em e.g.} for $\eta = -1/\epsilon$,
\be \label{zrendef}
 \zeta \to \overline\zeta := Z_\zeta(\epsilon) \, \zeta \,,
 \quad
 V_1 \to \overline V_1 := Z_\ssV (\epsilon) \, V_1  \,,
 \quad
 \chi_1 \to \overline \chi_1 := Z_\chi (\epsilon) \, \chi_1
 \quad \hbox{and so on} \,.
\ee
We define the parameters $Z_\ssV$, $Z_\zeta$ {\em etc.} so that $\overline \zeta$, $\overline V_1$ and the others remain finite. Since, as we show later, the integration constants like $\lambda_i^\pm$ are directly relatable to physical observables, the above matching conditions give us guidelines on how the various couplings renormalize. For instance, inspection of eq.~\pref{ceqn} shows that the product $\overline \zeta \, \overline \chi_1$ should remain finite, since it determines the physically measurable quantity $c$. Consequently
\be \label{zetachiZs}
 Z_\zeta(\epsilon) Z_\chi (\epsilon) = \hbox{finite} \,.
\ee

Next, the finiteness of $\zeta \chi_1$ together with the particular combination of matching conditions that sets $\lambda_3^-$ --- {\em i.e.} eqn. ~\pref{lambda3matching} --- shows that when $\eta = -1/\epsilon$ we must define
\be \label{eq:ZVexp}
 Z_\ssV = \frac{\overline V_1}{V_1} = e^{-\left[ \lambda_3^- + \frac32 \,
 (\lambda_2^- - \lambda_1) \right]/\epsilon} + \hbox{(finite)} \,,
\ee
in order to compensate for the divergent behaviour of $e^{\varphi + \beta - v}$.

Using this in the inflaton equation, eq.~\pref{inflaton-eom}, and keeping in mind that (see below) $H_0$ is a physical parameter, we find
\be
 H_0^2 \propto e^{-\( \lambda_1 - \lambda_2^- + \frac{2}{c} \lambda_3^-
 \) /\epsilon} \;  \frac{\zeta }{\chi_1} \,,
\ee
and so this, together with eq.~\pref{zetachiZs}, leads to $Z_\zeta (\epsilon) / Z_\chi(\epsilon) = \hbox{finite}$. If we absorb only the exponential dependence on $1/\epsilon$ into the renormalizations --- {\em e.g.} taking $\hbox{`finite'} = 0$ in eq.~\pref{eq:ZVexp} --- this implies
\ba
 Z_\zeta &=& e^{ - \frac12 \( \lambda_1 - \lambda_2^-
 + \frac{2}{c} \, \lambda_3^- \) / \epsilon} \nn\\
 Z_\chi &=& e^{\frac12 \( \lambda_1 - \lambda_2^-
 +\frac{2}{c} \, \lambda_3^- \)/\epsilon} \,.
\ea
Finally, the matching conditions involving $T_i$ are rendered finite by defining
\ba
 1 - \frac{\kappa^2 \overline T_i}{2\pi}  &:=& e^{- ( \lambda_2^- - \lambda_1 )/\epsilon} \( 1 - \frac{\kappa^2 T_i}{2\pi} \) + \hbox{(finite)} \,.
\ea
$\Phi_i$ does not require a divergent renormalization, as it appears as a finite quantity in the matching conditions.

\subsubsection*{Connection to physical properties}

Since the above section uses the finiteness of the bulk integration constants, $\lambda_i^\pm$, $H_0$, $c$ {\em etc.}, we pause here to relate these quantities more explicitly to physical observables. This ultimately is what allows us to infer the values taken by the finite renormalized parameters.

First, $c$ and $H_0$, directly determine the power of time with which the scale factor for the on-brane dimensions expand, and is thereby measurable through cosmological observations that determine $\dot a/a$, $\ddot a/a$ and so on.

Similarly, the volume of the extra dimensions is,
\be
 \cV_2 = \int \exd^2x \, \sqrt{g_2} =
 2\pi (H_0\tau)^c\tau^2\int_{-\infty}^\infty \exd\eta \,
 \exp\(\frac\cx2+\frac{3\cy}2 +\cz\),
\ee
and the proper distance between the branes is given by
\be
 L = (H_0\tau)^{c/2}\tau \int_{-\infty}^\infty
 \exd\eta \, \exp\( -\frac\cx4 +\frac{5\cy}4 + \frac\cz2 \) \,.
\ee
It is through relations such as these that physical quantities get related to the integration constants. In particular, convergence of these integrals implies conditions on the signs of the combinations $\lambda_1+4\lambda_2^\pm+2\lambda_3^\pm$ and $-\lambda_1+5\lambda_2^\pm+2\lambda_3^\pm$, all of which must be finite. The same is true of $\lambda_2$, which can be regarded as a function of the other two powers through the constraint \pref{eq:powersconstraint}.

Finally, the fluxes, $\Phi_s$ and $\Phi_i$, appear in the flux quantization condition and are directly related to a (finite) physical quantity: the magnetic charge of the branes. The renormalized tensions, $T_s$ and $T_i$, similarly enter into expressions for the deficit angle at the corresponding brane location.

\subsection{The 6D perspective in a nutshell}

Before turning to the view as seen by a 4D observer, this section first groups the main results obtained above when using the time-dependent matching conditions, eqs.~\pref{redmatching}, to relate the constants of the bulk scaling solution to the (renormalized) parameters in the source-brane actions, eqs.~\pref{eq:Sspec} and \pref{eq:Sinf}.

The physical couplings that we may specify on the inflaton brane are the renormalized quantities $V_1$, $\zeta $, $T_i$ and $\Phi_i$ (and we henceforth drop the overbar on renormalized quantities). On the spectator brane we similarly have $T_s$ and $\Phi_s$. We also get to specify  `initial conditions' for the on-brane inflaton: $\chi_0$ and $\chi_1$, as well as the integer, $n$, appearing in the flux-quantization condition. Of these, $\chi_0$ and $V_1$ only appear in the combination $\hat V_1 = V_1 \, e^{\zeta \chi_0}$, and so the value of $\hat V_1$ can be regarded as an initial condition for the inflaton rather than a choice for a brane coupling. Altogether these comprise 8 parameters: 5 brane couplings; 1 bulk flux integer; and 2 inflaton initial condition.

We now summarize the implications these parameters impose on the integration constants in the bulk, and identify any consistency conditions amongst the brane properties that must be satisfied in order to be able to interpolate between them using our assumed scaling bulk solution.

\subsubsection*{Time dependence}

First off, consistency of the scaling ansatz for the time dependence of all fields gives
\be \label{cvszetachi}
 c = -2 - \zeta \chi_1 \,.
\ee
Notice that this involves only the brane coupling $\zeta$ --- whose value determines the flatness of the inflaton potential --- and the inflaton initial condition, $\chi_1$. In particular, $c = -2$, corresponding to a de Sitter on-brane geometry, if either $\zeta$ or $\chi_1$ is chosen to vanish.

Next, we take the inflaton equation of motion on the brane to give the bulk parameter $H_0$ in terms of choices made on the inflationary brane:
\be
 H_0^2 = e^{-\frac12 (\cx_\infty^- - \cy_\infty^-)
 +\frac{\zeta \chi_1}{2+\zeta \chi_1} \, \cz_\infty^-}
 \left( \frac{\hat V_1}{3+2 \, \zeta \chi_1}
 \right) \frac{\zeta }{ \chi_1} \,.
\ee
Among other things, this shows that the choice $\chi_1 = 0$ does not satisfy the $\chi$ field equation unless $\zeta$ or $V_1$ vanish.

\subsubsection*{Consistency relations}

Consider next how the number of couplings on the branes restricts the other integration constants in the bulk.

Start with the spectator brane. Near the spectator brane we have $\lambda_3^+ = 0$ and
\be \label{lamsums}
 \lambda_1 = \lambda_2^+ = e^{\cy_\infty^+ - \cx_\infty^+}
 \(1-\frac{\kappa^2 T_s}{2\pi}\) \,,
\ee
as well as $\Phi_s = 0$. Specifying $T_s$ therefore imposes two relations among the four remaining independent bulk integration constants, $\lambda_1$, $\lambda_2^+$, $\cy_\infty^+$ and $\cz_\infty^+$, relevant to asymptotics near the spectator brane. We regard eq.~\pref{lamsums} as being used to determine the value of two of these, $\lambda_2^+$ and $\cy_\infty^+$ say.

Next we use the bulk equations of motion, eqs.~\pref{eq:chisoln} and \pref{bulkXY}, to integrate the bulk fields across to the inflaton brane. Starting from a specific choice for the fields and their $\eta$-derivatives at the spectator brane, this integration process leads to a unique result for the asymptotic behaviour at the inflaton brane. Given the 2-parameter set of solutions consistent with the spectator brane tension, integration of the bulk field equations should generate a 2-parameter subset of the parameters describing the near-inflaton-brane limit.

Now consider matching at the inflaton brane. The three asymptotic powers describing the near-brane limit for the inflaton brane can be expressed as
\ba \label{lamsumi}
 \lambda_1 &=& e^{\cz_\infty^- - \frac32 ( \cx_\infty^- -
  \cy_\infty^-)} \frac{\kappa^2 \hat V_1}{2\pi} \,
  \( \frac{\zeta \chi_1}{3+2 \, \zeta \chi_1} \)
  + e^{\cy_\infty^- - \cx_\infty^-} \( 1
  -\frac{\kappa^2 T_i}{2\pi} \)
  + \frac{3\kappa^2 H_0 q \, \Phi_i}{2\pi} \nn\\\nn\\
 \lambda_2^- &=& e^{ \cz_\infty^- - \frac32 ( \cx_\infty^-
 - \cy_\infty^-)} \frac{\kappa^2 \hat V_1}{2\pi}
 \( \frac{-6-3 \, \zeta \chi_1}{3+2 \, \zeta \chi_1} \)
 + e^{ \cy_\infty^- - \cx_\infty^-} \( 1 -
 \frac{\kappa^2 T_i}{2\pi} \)
 +\frac{\kappa^2 H_0 q\, \Phi_i}{2\pi}\\
\nn\\
 \lambda_3^- &=& e^{ \cz_\infty^- - \frac32 ( \cx_\infty^-
 - \cy_\infty^-)} \frac{\kappa^2 \hat V_1}{2\pi}
 \( \frac{6+3 \, \zeta \chi_1}{3+2 \, \zeta \chi_1} \) \,, \nn
\ea
which follow from three of the four matching conditions at the inflaton brane.\footnote{Recall that for time-independent systems there are 3 metric matching conditions -- $(tt)$, $(ij)$ and $(\theta \theta)$ -- plus that for the dilaton, $\phi$. The Hamiltonian constraint then imposes one relation amongst these three conditions, that can be regarded as implicitly fixing how the brane action depends on $g_{\theta\theta}$.} Notice that the constant $q$ appearing here can be regarded as being a function of the flux-quantization integer $n$ and the inflaton-brane flux coupling, $\Phi_i$:
\be
 q = \frac{4\pi g \lambda_1}{\kappa^2H_0[2\pi n -g\sum_b\Phi_b]}
  = \frac{4\pi g  \lambda_1 }{\kappa^2H_0[2\pi n -g\Phi_i]}\,.
\ee

The three parameters $\lambda_1$, $\lambda_2^-$ and $\lambda_3^-$ are not independent because they must satisfy the constraint, eq.~\pref{eq:powersconstraint},
\be \label{hamconst2}
 (\lambda_2^-)^2 - (\lambda_1)^2 = \frac43 \left( \frac{1+c+c^2}{c^2}
 \right) (\lambda_3^-)^2
 =\frac{12 + 12 \, \zeta \chi_1 +4(\zeta \chi_1)^2}{12
 +12\, \zeta \chi_1 +3(\zeta \chi_1)^2} \; (\lambda_3^-)^2 \,,
\ee
whose validity follows as a consequence of the field equations because the same constraint holds for the parameters, $\lambda_1$, $\lambda_2^+$ and $\lambda_3^+$, that control the bulk asymptotics near the spectator brane.

In principle, for a given set of inflaton-brane couplings we can regard two of eqs.~\pref{lamsumi} as fixing the remaining two free bulk parameters. The third condition does not over-determine these integration constants of the bulk, because the constraint, eq.~\pref{hamconst2}, is satisfied as an identity for all of the 2-parameter family of bulk solutions found by matching to the spectator brane. Consequently the third of eqs.~\pref{lamsumi} must be read as a constraint on one of the inflaton-brane properties. If we take this to be $\hat V_1$, say, then it can be interpreted as a restriction on the initial condition, $\chi_0$, in terms of the spectator-brane tension. This restriction is the consistency condition that is required if we wish to interpolate between the two branes using the assumed bulk scaling solution.

\subsubsection*{Inflationary choices}

In the end of the day we see that consistency with the bulk geometry does not preclude us from having sufficient freedom to adjust brane properties like $\zeta$ and $\chi_1$ to dial the parameters $c$ and $H_0$ freely. This shows that there is enough freedom in our assumed brane properties to allow treating these bulk parameters as independent quantities that can be freely adjusted.

In particular, we are free to choose the product $\zeta \chi_1$ to be sufficiently small and positive -- {\em c.f.} eq.~\pref{cvszetachi} -- to ensure an accelerated expansion: {\em i.e.} that $c$ is just slightly more negative than the de Sitter value of $-2$. This is the adjustment that is required to assure a `slow roll' within this model.

We also see that the time-dependence of the solution is such that the brane potential energy shrinks as the brane expands. That is, evaluated at the solution, eq.~\pref{chisoln},
\be \label{Vovertime}
 \Bigl. V_1 \, e^{\zeta \chi} \Bigr|_{\rm soln}
 = \hat V_1 \, | H_0 \tau
 |^{\zeta \chi_1}  = \hat V_1  \, \Bigl(|c+2|
  \, H_0 t  \Bigr)^{\zeta \chi_1/(2+c)}
  = \frac{\hat V_1}{ \zeta \chi_1
  H_0 t } \,.
\ee
This shows how inflation might end in this model. Suppose we take
\be
 V(\chi) = V_0 + V_1 \, e^{\zeta \chi} + V_2 \, e^{2 \, \zeta \chi}
 + \cdots \,,
\ee
where $V_1$ is chosen much larger than $V_0$ or the other $V_k$'s. Then if $\chi$ is initially chosen so that $V(\chi) \simeq V_1 \, e^{\zeta \chi}$ is dominated by the term linear in $e^{\zeta \chi}$, then the above scaling bulk solution can be consistent with the brane-bulk matching conditions. But eq.~\pref{Vovertime} shows that this term shrinks in size when evaluated at this solution (as also do the terms involving higher powers of $e^{\zeta \chi}$), until eventually the $V_0$ term dominates.

Once $V_0$ dominates the bulk scaling solution can no longer apply, plausibly also implying an end to the above accelerated expansion of the on-brane geometry. If $V(\chi) \simeq V_0$, then the inflaton brane effectively has a $\phi$-dependent tension, $T_{\rm eff} = T_i + V_0 \, e^{-\phi}$, which breaks the bulk scale invariance and so can lift the bulk's flat direction \cite{Cod2Matching, BBvN, susybranes} and change the dynamics of the bulk geometry.

Although this likely ends the inflationary evolution described above, it is unlikely in itself to provide a sufficiently graceful exit towards a successful Hot Big Bang epoch. Earlier calculations for maximally-symmetric branes show that such a tension leads to an effective potential (more about which below) proportional to $T_{\rm eff}' \propto - V_0 \, e^{-\phi}$, which points to a continued runaway along the would-be flat direction rather than a standard hot cosmology. We leave for further work the construction of a realistic transition from extra-dimensional inflation to later epochs, but expect that a good place to seek this interface is by modifying the assumption that $\Phi_s$ and/or $\Phi_i$ remain independent of $\phi$, since it is known \cite{TNCC} that when $\sum_b \Phi_b \propto e^{\phi}$ the low-energy scalar potential can act to stabilize $\phi$ at a minimum where the low-energy effective potential vanishes (classically).

\section{The view from 4D}

We now ask what the above dynamics looks like from the perspective of a 4D observer, as must be possible on general grounds within an effective theory in the limit when the Hubble scale, $H$, is much smaller than the KK scale. We can find the 4D description in this limit by explicitly compactifying the 6D theory. Our goal when doing so is to show how the low-energy 4D dynamics agrees with that of the explicit higher-dimensional solution, and to acquire a better intuition for how this inflationary model relates to more familiar 4D examples.

\subsection{The 4D action}

The simplest way to derive the functional form of the low-energy 4D action (at least at the classical level) is to use the classical scale invariance of the bulk field equations, since these are preserved by the choices we make for the branes --- at least during the inflationary epoch where $V \simeq V_1 \, e^{\zeta \chi}$.

Since this symmetry must therefore also be a property of the classical 4D action, there must exist a frame for which it can be written in the following scaling form:
\ba \label{4Deffaction}
 S_{\rm eff} &=& - \int \exd^4x \sqrt{ - \hat g_4}
 \; e^{-2 \varphi_4} \left[ \frac1{2\kappa_{4}^2}
 \hat g^{\mu\nu} \( \hat R_{\mu\nu} + Z_\varphi\,
 \pd_\mu \varphi_4 \pd_\nu\varphi_4 \) \right. \nn\\
 && \qquad\qquad\qquad\qquad\qquad\qquad
 \left. \phantom{\frac12}
  + f^2 \, \hat g^{\mu\nu}
 \pd_\mu \chi \pd_\nu \chi + U_\JF \( e^{\zeta \chi - \varphi_4} \)
 \right] \\
 &=& - \int \exd^4x \sqrt{ - {\bf g}_4}
 \; \left[ \frac1{2\kappa_{4}^2}
 {\bf g}^{\mu\nu} \( {\bf R}_{\mu\nu} + (6 + Z_\varphi) \,
 \pd_\mu \varphi_4 \pd_\nu\varphi_4 \) \right. \nn\\
 && \qquad\qquad\qquad\qquad\qquad\qquad
 \left. \phantom{\frac12}
  + f^2 \, {\bf g}^{\mu\nu}
 \pd_\mu \chi \pd_\nu \chi + e^{2 \varphi_4}
 U_\JF \( e^{\zeta \chi - \varphi_4} \)
 \right] \,, \nn
\ea
where $\varphi_4$ denotes the 4D field corresponding to the flat direction of the bulk supergravity and $\chi$ is the 4D field descending from the brane-localized inflaton. The second version gives the action in the 4D Einstein frame, whose metric is defined by the Weyl transformation:
\be
 {\bf g}_{\mu\nu} = e^{-2\varphi_4} \hat g_{\mu\nu} \,.
\ee
The potential, $U_\JF$, is an a-priori arbitrary function of the scale-invariant combination $e^{\zeta \chi - \varphi_4}$, whose functional form is not dictated purely on grounds of scale invariance.

The detailed form of $U_\JF$ and the values of the constants $\kappa_4$, $Z_\varphi$ and $f$, are calculable in terms of the microscopic parameters of the 6D theory by dimensional reduction. As shown in detail in Appendix \ref{app:dimred}, we find $Z_\varphi = -4$,
\ba \label{kappaJFandfmatching}
 \frac{1}{2 \kappa_4^2} &=&  \int \exd \theta \exd \eta \;
 \frac{e^{-\omega + 3\alpha + \beta + v}}{2 \kappa^2 H_0^2}
 = \frac\pi{ \kappa^2 H_0^2} \int \exd \eta
 \; e^{2\cy-2\cz/c} \nn\\
 &=& \frac\pi{ \kappa^2 H_0^2} \int \exd \eta \; \frac{\cz''}{3c}
 = -\frac{\pi\lambda_3^-}{H_0^2\kappa^2c} \nn\\
 f^{2} &=& e^{-\cx_\infty^- +\cy_\infty^- - \frac2c\cz_\infty^-}
 \left( \frac{23-2c}{28+8c} \right) \,,
\ea
while the potential becomes
\be \label{VEFmatching}
 V_\EF := e^{2\varphi_4} \, U_\JF =
 - C e^{2\varphi_4} + D e^{\zeta\chi + \varphi_4} \,,
\ee
with the constants $C$ and $D$ evaluating to
\ba \label{CDmatching}
 C &=& \frac54 \, q  H_0 \Phi_i -  e^{-\cx_\infty^- + \cy_\infty^-}
 \(\frac{2\pi}{\kappa^2} -  T_i \)
 - e^{-\cx_\infty^+ + \cy_\infty^+}  T_s \nn\\
 D &=& \frac54 e^{-\frac32 (\cx_\infty^- - \cy_\infty^-)
 + \cz_\infty^-} V_1\,.
\ea
In the regime of interest, with $\kappa^2 T_i/2\pi \ll 1$ and $\kappa^2 T_s /2\pi \ll 1$ and $V_1 > 0$, both $C$ and $D$ are positive. The unboundedness from below of $V_\EF$ as $\varphi_4 \to \infty$ is only an apparent problem, since the domain of validity of the semiclassical calculations performed here relies on the bulk weak-coupling condition, $e^{\varphi_4} \ll 1$.

\subsection*{4D dynamics}

The classical field equations obtained using this 4D effective action consist of the following scalar equations,
\ba
 \frac{2}{\kappa_{4}^2} \, \Box\varphi_4 &=&
 -2C \, e^{2\varphi_4} + D \, e^{\zeta \chi + \varphi_4} \nn\\
 2{f^2} \, \Box\chi &=& \zeta D \, e^{\zeta \chi + \varphi_4} \,,
\ea
and the trace-reversed Einstein equations
\be
 {\bf R}_{\mu\nu} + 2 \, \pd_\mu \varphi_4 \pd_\nu\varphi_4
 + {2\kappa_{4}^2}{f^2} \, \pd_\mu \chi \pd_\nu \chi
 +\kappa_{4}^2 V_{\EF} \, {\bf g}_{\mu\nu} = 0 \,.
\ee

This system admits scaling solutions, with all functions varying as a power of time,
\ba \label{4Dpowerlaw}
 {\bf g}_{\mu\nu} &=& (H_0\tau)^{2+2c}
 \( \eta_{\mu\nu} \, \exd x^\mu \exd x^\nu \) \nn\\
 e^{\varphi_4} &=& e^{\varphi_{40}} (H_0\tau)^{-2-c} \nn\\
 e^{\zeta\chi} &=& e^{\zeta \chi_0} (H_0\tau)^{\zeta\chi_1}
 = e^{\zeta \chi_0} (H_0\tau)^{-2-c}\,.
\ea
Notice that the consistency of the field equations with the power-law time-dependence requires $\zeta \chi_1=-2-c$, just like in six dimensions ({\em c.f.} eq.~\pref{ceqn}). With this, the scalar equations of motion are
\ba
 \frac{2}{\kappa_{4}^2} \,H_0^2 (2c^2+5c+2) &=&
 -2 C \, e^{2 \varphi_{40}} + D \, e^{\zeta \chi_0 + \varphi_{40}} \nn\\
 -2 (2c+1) {H_0^2 \, \chi_1}{f^2} &=&
 \zeta D \, e^{\zeta \chi_0 + \varphi_{40}} \,,
\ea
and the Einstein equations become
\ba
 \frac{H_0^2}{\kappa_{4}^2} \( 2c^2+5c+5 \)
 + {2H_0^2 \, \chi_1^2}{f^2}
 &=& - C\, e^{2 \varphi_{40}}
 +D \, e^{\zeta \chi_0 + \varphi_{40}} \nn\\
 \frac{H_0^2}{\kappa_{4}^2} (2c^2+3c+1) &=&
 -C\, e^{2 \varphi_{40}}
 +D\, e^{\zeta \chi_0 + \varphi_{40}} \,.
\ea
These four equations are to be solved for the three variables $\chi_0$, $\chi_1$ and $\varphi_{40}$ appearing in the power-law ansatz, eqs.~\pref{4Dpowerlaw}. This is not an over-determined problem because the four equations are not independent (a linear combination of the two scalar equations gives the second Einstein equation).

Subtracting the two Einstein equations yields
\be
 {\chi_1^2}{f^2} = - \frac{2+c}{\kappa_{4}^2}
 = \frac{\zeta\chi_1}{\kappa_{4}^2} \,,
\ee
and so discarding the trivial solution, $\chi_1=0$, we find
\be \label{eq:chi1soln}
 {\chi_1} = \frac{\zeta}{\kappa_{4}^2 f^2} \,.
\ee
Next, dividing the two scalar equations gives the relation
\be \label{solntoscalareqn}
 -\frac{2c^2+5c+2}{(2c + 1) \kappa_4^2 f^2 \chi_1} =
 -\frac{c+2}{\zeta} = \frac{1}{\zeta} \(1
 - \frac{2C}{D} \, e^{\varphi_{40} - \zeta \chi_0 } \) \,,
\ee
where the first equality uses eq.~\pref{eq:chi1soln}. Combining eqs.~\pref{ceqn}, \pref{eq:chi1soln} and \pref{solntoscalareqn} finally gives
\be \label{chivsCD}
 \frac{\zeta^2}{\kappa_{4}^2 f^2} = 1 - \frac{2C}{D} \, e^{\varphi_{40}
 - \zeta \chi_0 } \,.
\ee
This last equation shows that the scaling ansatz is only consistent with the field equations if $\chi_0$ is chosen appropriately, in agreement with what was found by matching between branes in the 6D perspective. It also shows, in particular, that $\zeta \chi_1$ can be dialed to be small and positive by suitably adjusting the scale-invariant (and time-independent) quantity $\varphi_4 - \zeta \chi$ so that the right-hand side of eq.~\pref{chivsCD} is sufficiently small and positive. This is not inconsistent with the microscopic choices made for the branes because the ratio $C/D$ is positive.

The upshot is this: the above relations precisely reproduce the counting of parameters and the properties of the solutions of the full 6D theory, once the low-energy parameters $C$, $D$, $\kappa_4$ and $f$ are traded for the underlying brane properties, using eqs.~\pref{kappaJFandfmatching} and \pref{CDmatching}.

\subsection{The 4D inflationary model}

The 4D effective description also gives more intuition of the nature of the inflationary model, and why the scalar evolution can be made slow.

Notice that the action, eq.~\pref{4Deffaction}, shows that the scalar target space is flat in the Einstein frame. Consequently, the slow-roll parameters are controlled completely by the Einstein-frame potential, eq.~\pref{VEFmatching}. In particular,
\ba
 \varepsilon_\varphi &:=& \left(
 \frac{1}{V_\EF} \; \frac{\partial V_\EF}{\partial
 \varphi_4} \right)^2 = \left(
 \frac{ - 2 +  (D/C) e^{\zeta\chi
 - \varphi_4}}{ - 1 + (D/C) e^{\zeta\chi
 - \varphi_4}} \right)^2 \nn\\
 \varepsilon_\chi &:=& \frac{1}{\kappa_4^2 f^2}
 \left( \frac{1}{V_\EF} \; \frac{\partial V_\EF}{\partial \chi}
 \right)^2 = \frac{\zeta^2}{\kappa_4^2 f^2} \left(
 \frac{ (D/C) e^{\zeta\chi - \varphi_4} }{
 - 1 + (D/C) e^{\zeta\chi - \varphi_4}}
 \right)^2 \,.
\ea
This shows that there are two conditions required for $V_\EF$ to have sufficiently small first derivatives for slow-roll inflation. First, $\varepsilon_\chi \ll 1$ requires $\zeta^2 \ll \kappa_4^2 f^2$, in agreement with the 6D condition $\zeta \chi_1 \ll 1$ once eq.~\pref{eq:chi1soln} is used. Second, $\varepsilon_\varphi \ll 1$ is generically {\em not} true, but can be made to be true through a judicious choice of initial conditions for $\zeta \chi - \varphi_4$: $(D/C) \, e^{\zeta \chi - \varphi_4} = 2 + \cO(\zeta \chi_1)$, in agreement with eq.~\pref{chivsCD}. Notice that in this case $\varepsilon_\chi \simeq \cO[ \zeta \chi_1 ]$ while $\varepsilon_\varphi \simeq \cO[(\zeta \chi_1)^2] \ll \varepsilon_\chi$.

Next, consider the second derivatives of $V_\EF$:
\ba
 \eta_{\varphi\varphi} &:=& \left(
 \frac{1}{V_\EF} \; \frac{\partial^2 V_\EF}{\partial
 \varphi_4^2} \right) =
 \frac{ - 4 +  (D/C) e^{\zeta\chi
 - \varphi_4}}{ - 1 + (D/C) e^{\zeta\chi
 - \varphi_4}} \simeq -2 + \cO(\zeta \chi_1) \nn\\
 \eta_{\varphi\chi} &:=& \frac{1}{\kappa_4 f}
 \left( \frac{1}{V_\EF} \; \frac{\partial^2 V_\EF}{
 \partial \varphi_4 \partial \chi}
 \right) = \frac{\zeta}{\kappa_4 f} \left(
 \frac{ (D/C) e^{\zeta\chi - \varphi_4} }{
 - 1 + (D/C) e^{\zeta\chi - \varphi_4}} \right)
 \simeq \frac{2\,\zeta}{\kappa_4 f}  + \cO(\zeta \chi_1)\\
  \eta_{\chi\chi} &:=& \frac{1}{\kappa_4^2 f^2}
 \left( \frac{1}{V_\EF} \; \frac{\partial^2 V_\EF}{
 \partial \chi^2}
 \right) = \frac{\zeta^2}{\kappa_4^2 f^2} \left(
 \frac{ (D/C) e^{\zeta\chi - \varphi_4} }{
 - 1 + (D/C) e^{\zeta\chi - \varphi_4}} \right)
 \simeq \frac{2\,\zeta^2}{\kappa_4^2 f^2} + \cO(\zeta \chi_1) \,,\nn
\ea
where the last, approximate, equality in each line uses eq.~\pref{chivsCD}.

\FIGURE[ht]{
\epsfig{file=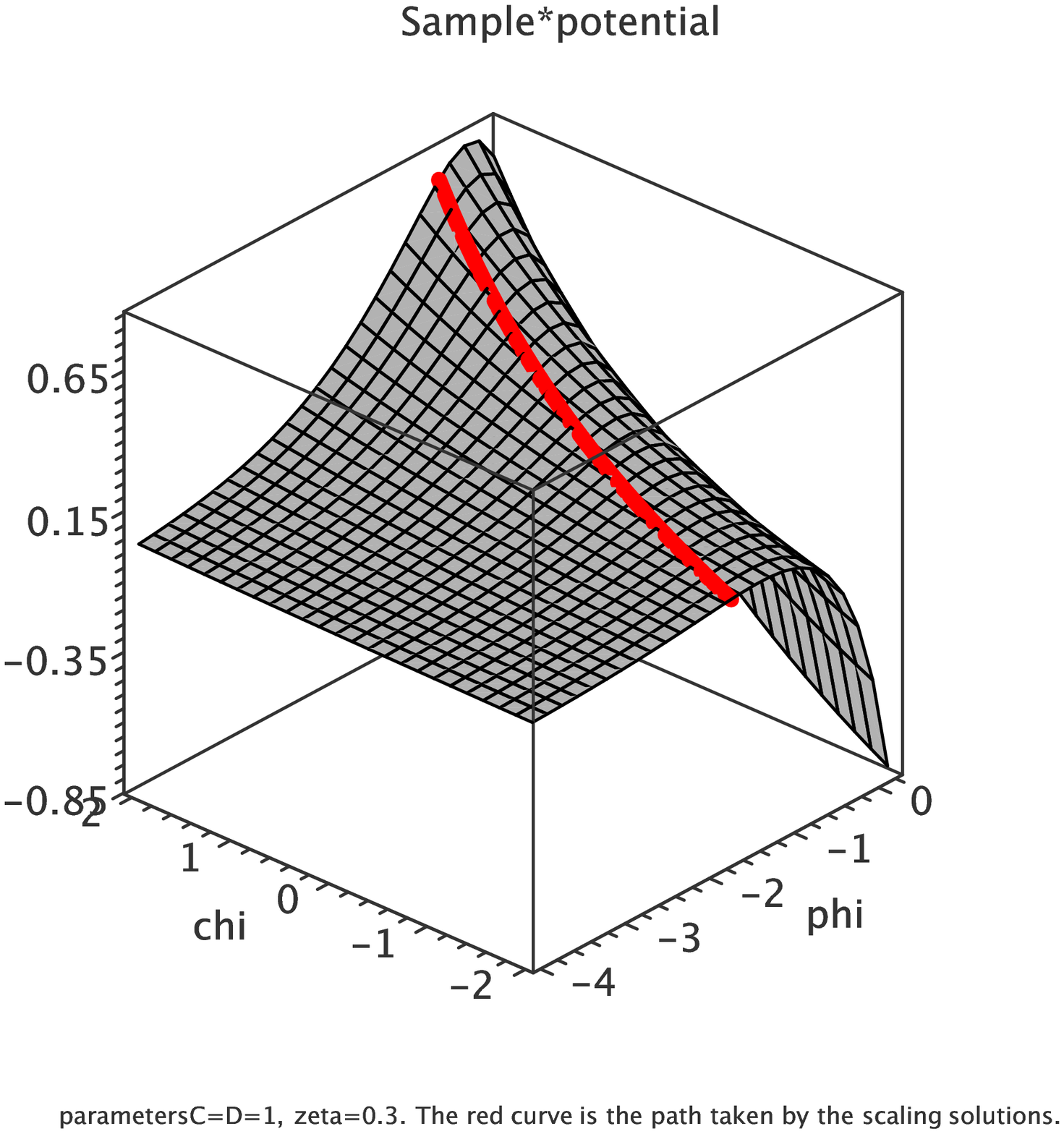,angle=0,width=0.45\hsize}
\caption{Sample potential evaluated for $C=D=1$ and $\zeta=0.3$. The red line denotes the path taken by the scaling solutions.
} \label{fig:potential} }

Notice that $\eta_{\varphi\varphi}$ is not itself small, even when $\zeta \ll \kappa_4 f$ and eq.~\pref{chivsCD} is satisfied. However, in the field-space direction defined by $\vec n := \vec\varepsilon/|\vec \varepsilon|$ we have $n_\chi \simeq \cO(1)$ and $n_\varphi \simeq \cO(\zeta \chi_1)$ and so
\be
 \eta_{ab} n^a n^b = \cO( \zeta \chi_1) = \cO \left(
 \frac{ \zeta^2}{ \kappa_4^2 f^2} \right) \ll 1 \,.
\ee
Because $\eta_{\varphi\varphi}$ is negative and not small, slow roll is achieved only by choosing initial conditions to lie sufficiently close to the top of a ridge, with initial velocities chosen to be roughly parallel to the ridge (see Fig.~\ref{fig:potential}). For single-field 4D models such an adjustment is unstable against de Sitter fluctuations of the inflaton field, and although more difficult to compute in the higher-dimensional theory, the low-energy 4D potential suggests that similar considerations are likely also to be true here.

\section{Conclusions}

In a nutshell, the previous sections describe a family of --- previously known \cite{scaling solutions} --- exact, explicit, time-dependent solutions to the field equations of 6D supergravity in the presence of two space-filling, positive-tension source branes. The solutions describe both the cosmological evolution of the on-brane geometry and the change with time of the extra-dimensional geometry transverse to the branes. These solutions have explicitly compact extra dimensions, with all but one modulus stabilized using an explicit flux-stabilization mechanism. The time evolution describes the dynamics of the one remaining would-be modulus of the bulk geometry to the back-reaction of the source branes.

\subsection{Bugs and features}

The new feature added in this paper is to identify a choice for the dynamics of a brane-localized scalar field whose evolution is consistent with the bulk evolution, and so can be interpreted as the underlying dynamics that gives rise to the bulk evolution. In order to find this choice for the brane physics we set up and solve the codimension-two matching problem for time-dependent brane geometries, extending earlier analyses \cite{Cod2Matching, BBvN, susybranes} of these matching conditions for systems with maximally symmetric on-brane geometries.

We also find the 4D theory that describes this system in the limit of slow evolution, where a low-energy effective field theory should apply. The low-energy theory turns out to be a simple scalar-tensor system involving two scalar fields in 4 dimensions: one corresponding to the brane-localized mode and one corresponding to the would-be flat direction of the bulk geometry. We verify that the 4D system has time-dependent solutions that reproduce those of the full 6D equations (as they must).

In particular, we identify a region of parameter space that describes an inflationary regime, including a limit for which the on-brane geometry is de Sitter. (The de Sitter solution is not a new one \cite{6DdS}, and evades the various no-go theorems \cite{dSnogo} because the near-brane behavior of the bulk fields dictated by the brane-bulk matching does not satisfy a smoothness assumption --- `compactness' --- that these theorems make.) For parameters near the de Sitter limit, the evolution is accelerated and takes a power-law slow-roll form, $a(t) \propto t^p$ with $p > 1$. (The de Sitter solution is obtained in the limit $p \to \infty$.) From the point of view of the low-energy 4D theory, the de Sitter solution corresponds to sitting at the top of a ridge, and the scaling solutions describe motion near to and roughly parallel with this ridge. Experience with the 4D potential suggests that the initial conditions required to obtain inflation in this model are likely to require careful tuning.

{}From the 4D perspective, the inflationary scenario resembles old models of extended inflation \cite{ExtInf}, for which accelerated power-law expansion is found to arise when Brans-Dicke theory is coupled to matter having an equation of state $w = -1$. Having a Brans-Dicke connection is perhaps not too surprising, despite earlier difficulties finding extended inflation within a higher-dimensional context. Part of what is new here relative to early work is the scale invariance of the bulk supergravity that is not present, for example, in non-supersymmetric 6D constructions \cite{ExtInfKK}. Another new feature is brane-localized matter, which was not present in early searches within string theory \cite{ExtInfStr}. Brans-Dicke-like theories arise fairly generically in the low-energy limit of the 6D supergravity of interest here because back-reaction tends to ensure that the bulk dilaton, $\varphi_4$, couples to brane-localized brane matter in this way \cite{BulkAxions, susybranes}.

For cosmological applications it is interesting that the 4D limit of the higher-dimensional system is not {\em exactly} a Brans-Dicke theory coupled to matter. It differs by having a scalar potential (rather than a matter cosmological constant), that is calculable from the properties of the underlying branes. It also differs by being `quasi-Brans Dicke', in that the scalar-matter coupling tends to itself depend on the Brans-Dicke field, $\varphi_4$. Both of these features are potentially attractive for applications because successful cosmology usually requires the Brans-Dicke coupling to be relatively large during inflation compared with the largest values allowed by present-day solar-system constraints \cite{ExtInfProb}. Having both field-dependent couplings and a scalar potential can allow these properties to be reconciled, by having the potential drive the scalar at late times to a value for which the coupling is small. (See, for instance, \cite{6Dquint} for a sample cosmology which uses this mechanism in a related example.)

A noteworthy feature of the inflationary geometries is that the extra dimensions are not static (although they become static in the strict de Sitter limit). Instead they expand with $r(t) \propto \sqrt t$, while the scale factor of the on-brane directions expands even faster, $a(t) \propto t^p$ with $p > 1$. As a result the Kaluza-Klein mass scale shrinks, as does the higher-dimensional gravity mass scale (measured in 4D Planck units), during the inflationary expansion.

If embedded into a full inflationary picture, including the physics of the late-epoch Hot Big Bang, such an inflationary scenario can have several attractive properties. First, the relative expansion rates of the various dimensions might ultimately explain why the four on-brane dimensions are much larger than the others. It might also explain why two internal dimensions might be bigger than any others, if it were embedded into a 10-dimensional geometry with the `other' 4 dimensions stabilized.

A second attractive feature is the disconnect that this scenario offers between the gravity scale during inflation and the gravity scale in the present-day universe.\footnote{In this our model is similar in spirit to ref.~\cite{VolInf}.} Inflationary models such as these can allow the current gravity scale to be low (in the multi-TeV range in extreme cases), and yet remain consistent with the observational successes of generating primordial fluctuations at much higher scales. Inflationary models like this might also point to a way out of many of the usual cosmological problems faced by low gravity-scale models \cite{ADD, MSLED}, such as a potentially dangerous oversupply of primordial KK modes.

\subsection{Outstanding issues}

The model presented here represents only the first steps down the road towards a realistic inflationary model along these lines, however, with a number of issues remaining to be addressed. Perhaps the most important of these are related to stability and to ending inflation and the transition to the later Hot Big Bang cosmology. Besides identifying the Standard Model sector and how it becomes reheated, it is also a challenge to identify why the cosmic expansion ends and why the present-day universe remains four-dimensional and yet is so close to flat.

What is intriguing from this point of view is the great promise that the same 6D supergravity used here also has for addressing some of these late-universe issues \cite{TNCC}, especially for the effective cosmological constant of the present-day epoch. In particular, these 6D theories generically lead to scalar-tensor theories at very low energies,\footnote{Remarkably, the same mechanism that can make the vacuum energy naturally small in 6D supergravity also protects this scalar's mass to be very light \cite{6Dquint, TNCC, susybranes}.} and so predict a quintessence-like Dark Energy \cite{6Dquint}. Successfully grafting the inflationary scenario described here onto this late-time cosmology remains unfinished, yet might provide a natural theory of initial conditions for the quintessence field as arising as a consequence of an earlier inflationary period (see \cite{QInf} for some other approaches to this problem, and \cite{DERev} for a more comprehensive review).

Other outstanding issues ask whether (and if so, how) the extra dimensions help with the problems of many 4D inflationary models: initial-condition problems, fine-tuning and naturalness issues, and so on. Since some of these questions involve `Planck slop' coming from the UV completion \cite{SIreviews}, a helpful step in this direction might be to identify a stringy provenance for the 6D gauged chiral supergravity studied here \cite{CP}.

Another interesting direction asks about the existence and properties of cosmological solutions that explore the properties of the extra dimensions more vigorously than is done by the model considered here. That is, although our model here solves the full higher-dimensional field equations, it is only the volume modulus of the extra-dimensional geometry that evolves with time, with all of the other KK modes not changing. Although our calculation shows that this is consistent with the full equations of motion, even for Hubble scales larger than the KK scale, it is probably not representative of the general case when $H > m_\KK$. More generally one expects other KK modes to become excited by the evolution, allowing a richer and more complex evolution.

There remains much to do.

\section*{Acknowledgements}

We thank Allan Bayntun and Fernando Quevedo for helpful discussions. Our research was supported in part by funds from the Natural Sciences and Engineering Research Council (NSERC) of Canada. Research at the Perimeter Institute is supported in part by the Government of Canada through Industry Canada, and by the Province of Ontario through the Ministry of Research and Information (MRI).

\appendix

\section{Asymptotic near-brane forms}
\label{app:AsForm}

This section establishes that the functions $\cy$ and $\cz$ must asymptote at large $|\eta|$ to linear functions of $\eta$. This conclusion follows from the conditions $e^\beta \to 0$ and $e^{\beta + v} \to 0$, that are argued in the main text to be consequences of the condition that the source branes be codimension-two objects within a finite-volume compactified bulk.

In terms of the functions $\cx$, $\cy$, $\cz$, the conditions $e^\beta \to 0$ and $e^{\beta + v} \to 0$ mean
\be
 \frac{3\cx}{4} + \frac{\cy}{4} + \frac{\cz}{2} \to -\infty
 \quad \hbox{and} \quad
 \frac\cx2 + \frac{3\cy}2 + \cz \to -\infty \,.
\ee
Since eq.~\pref{eq:chisoln} shows $\cx \to -\infty$ linearly with $|\eta|$ at both branes, this means that the two combinations $\cy + 2\cz$ and $3\cy+2\cz$, if positive, can grow at worst linearly in $|\eta|$ as $\eta \to \pm \infty$. This implies that the second derivatives of those combinations are bounded by
\ba
 \lim_{\eta\to\infty}(\cy''+2\cz'')\leq0\nn\\
 \lim_{\eta\to\infty}(3\cy''+2\cz'')\leq0\,.
\ea
Explicit expressions for those derivatives state
\ba
\label{difeqn}
 \cy''+2\cz'' &=& e^{2\cy}\left[H_0^2 \, (4+10c+4c^2) \, e^{-2\cz/c}-\frac{4g_\ssR^2}{\kappa^2}\right]\nn\\
 3\cy''+2\cz'' &=& e^{2\cy}\left[H_0^2 \, (12+18c+12c^2) \, e^{-2\cz/c}-\frac{12g_\ssR^2}{\kappa^2}\right] \,.
\ea

Now there are a few cases to consider. First, suppose neither $\cy$ nor $\cz$ remains finite. Since the coefficient in front of $\exp(-2\cz/c)$ is positive, this term has to vanish: If instead it diverges, it dominates the terms in brackets which turns the derivatives positive as $\eta\to\infty$. Now also $\cy\to-\infty$, because if we assume $\cy\to\infty$ we find $\cy''\-4g_\ssR^2 e^{2\cy}/\kappa^2\to-\infty$, inconsistent with $\cy\to\infty$. Hence, if both $\cy$ and $\cz$ diverge the two exponentials vanish and therefor both $\cy''$ and $\cz''$ vanish.

If $\cy$ approaches a finite value, then both $\cy'\to0$ and $\cy''\to0$. Since in that case both eqns ~\pref{difeqn} become $\cz''=...$ as $\eta\to\infty$, the terms in square brackets need to equal. They don't unless $\cz$ approaches a finite value, which means $\cz'\to0$ and $\cz''\to0$.

Finally, assume $\cz$ stays finite but $\cy\to\pm\infty$. This again implies $\cz''\to0$, so if we now compare the two eqns ~\pref{difeqn} the terms in brackets in the second equation has to be three times the term in the first equation, or the whole right hand side has to vanish for both. The two terms are equal only if $c=0$, so this is the last possible exception to linear behaviour: $c=0$, $\cz$ stays finite and $\cy\to\infty$ in some less fast than linear way. In this case, the differential equation for $\cy$ approaches
\be
 \cy''= K e^{2\cy}\,,
\ee

If $K$ is positive, the solutions diverge much faster than linear. If $K$ is negative, the solution approaches a function of the form
\be
 e^\cy\approx\cosh^-(\lambda_2(\eta-\eta_2))\,,
\ee
which indeed has a limit of $\cy$ being linear. This covers all cases, so demanding finite volume and codimension-2 branes implies linear limits for $\cy$ and $\cz$. Finally we point out that if $\lambda_3=0$ at a given brane, and $c\neq0$, the LHS of ~\pref{difeqn} has to vanish in the limit at that brane, and this is only possible if $e^{2\cy}$ vanishis ($\cy\to-\infty$). In our conventions, this means $\lambda_2>0$ at that brane.

\section{Flux quantization with the brane flux}
\label{app:FluxQ}

This appendix reviews the derivation \cite{BulkAxions, susybranes} of the form of the flux-quantization condition in the case where flux can be localized on the branes.

We follow the derivation in \cite{susybranes}, only slightly modifying the discussion due to our different choice of coordinates in the current context. The brane action we regularize with some function $s(\eta)$ which is zero for $\eta>\eta_b$, and will eventually take $\eta_b\to-\infty$. We normalize the weighting function as
\be
 \int_0^{2\pi} \exd\theta \int_{-\infty}^{\eta_b}
 \exd\eta\sqrt{g_2}s(\eta)=1 \,.
\ee

The full action for the gauge field is
\be
 S=-\int \exd^6x \sqrt{-g} \left[\frac{e^{-\phi}}4\cF_{\ssM\ssN}\cF^{\ssM\ssN} + \frac{e^{\phi}}2\Phi\epsilon^{mn}\cF_{mn} \, s(\eta)\right]\,,
\ee
with equation of motion (given our symmetry ans\"atze)
\be
 \pd_\eta\(\sqrt{-g}e^{-\phi}g^{\eta\eta}g^{\theta\theta}\pd_\eta \cA_\theta - \Phi e^{-\phi} \sqrt{-\gamma}s(\eta)\)=0\,.
\ee
This can be integrated once to given
\be
 \pd_\eta\cA_\theta = \frac{\sqrt{g_2}}{\sqrt{-\gamma}}
 \mathcal Q e^\phi + \Phi \sqrt{g_2} s(\eta)\,.
\ee

The value of the gauge potential near the brane, assuming it vanishes at $-\infty$, is then
\ba
 2\pi\cA_\theta(\eta_b)&=&\int_0^{2\pi}\exd\theta\int_{-\infty}^{\eta_b}\exd\eta \pd_\eta\cA_\theta\nn\\
 &=& \Phi +2\pi\mathcal Q\int_{-\infty}^{\eta_b} e^\phi\frac{\sqrt{g_2}}{\sqrt{-\gamma}}\nn\\
 &=& \Phi +2\pi\mathcal Q\frac{H_0^2}{(H_0\tau)^2}\int_{-\infty}^{\eta_b} e^{\varphi+v+\beta-\omega-\alpha}\nn\\
 &=& \Phi +2\pi\mathcal Q\frac{H_0^2}{(H_0\tau)^2}\int_{-\infty}^{\eta_b} e^{2\cx}\nn\\
 &\approx& \Phi +2\pi\mathcal Q\frac{H_0^2}{(H_0\tau)^2}\frac{e^{2\lambda_1\eta_b}}{2\lambda_1}
\ea
{}From the last equation we see that as we take $\eta_b\to-\infty$ this approaches $\Phi$, corresponding to our use in the main text.

\section{Derivation of time-dependent codimension-two matching}
\label{matchingderivation}

This appendix derives the form of the codimension-two matching conditions used in the main text for time-dependent problems. It does so by following the strategy used in refs.~\cite{Cod2Matching} in the maximally symmetric case, wherein the codimension-two brane is modeled as a very small cylindrical codimension-one brane whose interior bulk field configuration is smooth.

At first sight this may seem to be an arbitrary construction, which would not be expected to capture the matching for different types of codimension-two objects. However, we follow the spirit of \cite{Cod2Matching} which argues that the matching conditions are very general, and capture the influence of the branes on the geometry far from the brane for {\em any} codimension-two brane, provided only that this brane is axially symmetric. The generality of this result is similar to the generality of the multipole expansion for electrodynamics: since the far-field behaviour of an electromagnetic field is controlled by the leading few multipole moments, it is accurately captured by {\em any} charge distribution that shares these few moments with the real source of interest. This argument is tested in ref.~\cite{BBvN}, by applying it to the case of $D7$ branes in 10D Type IIB supergravity, and found to work extremely well.

The virtue of trading the codimension-two brane for a codimension-one cylinder is that it allows the use of the familiar Israel junction conditions \cite{IJC} to infer the near-brane geometry external to the codimension-two object. For this reason we first set up the geometric description of the extrinsic geometry of the small codimension-two cylinder in a way that includes nontrivial time-dependence.

We here start our derivation by imagining excising the source branes from the bulk geometry, and asking about the variation of the action on the boundary of this excision. We regard this boundary as a codimension-one brane, with an interior filled using a smooth geometry. The codimension-two brane action can be regarded as the dimensional reduction of the action on this fictitious codimension-one brane.

The boundary part of the variation have two parts to it: First, there is a part coming from the bulk action through stokes' theorem. Second, there is the direct contribution from varying the boundary action. The next sections calculate the contribution from the bulk action for branes that are located at constant coordinate $\rho$, but without assuming that $\rho$ is perpendicular to the brane, or even that the off diagonal metric components $g_{\rho m}$ vanish. To do this we write the action in the ADM decomposition, but with $\rho$ playing the role of time.

\subsection*{Conventions and notation for ADM decomposition}

We describe spacetime as surfaces of constant $\rho$. As in the main text, we use capital indices $M$, $N$, for the full spacetime, and small indices $m$, $n$, for coordinates on the slice (ie, all except for $\rho$) The unit normal to those surfaces is denoted by $N_\ssM$, and has components
\be
N_\rho=\frac1{\sqrt{g^{\rho\rho}}}\qquad\qquad N_m=0\,.
\ee
Since we are not assuming a diagonal metric, this does not imply $N^m=0$. The projection operator on the slices is
\be
\proj^\ssM_\ssN=\delta^\ssM_\ssN-N^\ssM N_\ssN\,.
\ee
The extrinsic curvature of the slices is defined as
\be
K_{\ssM\ssN}=\proj_\ssM^\ssP\proj_\ssN^\ssQ\nabla_\ssP N_\ssQ\,,
\ee
and the intrinsic curvature is defined as follows: For any vector $V^\ssM$ in the surface, ie. $N_\ssM V^\ssM=0$, we define the covariant derivative $D_\ssM$ as the full covariant derivative projected back to the surface:
\be
D_\ssM V^\ssN = \proj^\ssP_\ssM \proj^\ssQ_\ssN \nabla_{\ssP} V^\ssQ
\ee
and similar for any tensor. The intrinsic curvature is defined with respect to this covariant derivative as
\be
\hat R^\ssM_{\ \ssN\ssP\ssQ}V^\ssQ=[D_\ssN,D_\ssP]V^\ssM
\ee

\subsection*{The Gauss-Codazzi equations and the action}
In this section we show that the Einstein-Hilbert action with the Gibbons-Hawking boundary term combine to
\be
S_{EH}+S_{GH}=-\int\exd^6x\sqrt{-g}\frac1{2\kappa^2}\(\hat R+K_{\ssM\ssN}K^{\ssM\ssN}-K^2\)\,.
\ee
Our starting point is the Gauss-Codazzi equation that states
\be
(\proj^4 R)_{\ssM\ssN\ssR\ssT}=\hat R_{\ssM\ssN\ssR\ssT} - K_{\ssM\ssT}K_{\ssN\ssR} + K_{\ssM\ssR}K_{\ssN\ssT}\,,
\ee
where $\proj^4$ is shorthand for projecting all the indices back to the slice. In order to write the Einstein-Hilbert action in terms of the intrinsic and extrinsic curvatures, we use that
\ba
R&=& g^{\ssM\ssR}g^{\ssN\ssT}R_{\ssM\ssN\ssR\ssT}\nn\\
&=&(\proj^{\ssM\ssR}+N^\ssM N^\ssR)(\proj^{\ssN\ssT}+N^\ssN N^\ssT)R_{\ssM\ssN\ssR\ssT}\nn\\
&=&\proj^{\ssM\ssR}\proj^{\ssN\ssT}R_{\ssM\ssN\ssR\ssT}+ 2\proj^{\ssM\ssR}N^\ssN N^\ssT R_{\ssM\ssN\ssR\ssT}\,,
\ea
which uses the symmetries of the Riemann tensor. Using the Gauss-Codazzi equation, we can write the first term as
\be
\proj^{\ssM\ssR}\proj^{\ssN\ssT}R_{\ssM\ssN\ssR\ssT}=\hat R - K_\ssM^\ssN K_\ssN^\ssM+K^2
\ee
The second term we can rewrite by relating it to derivatives of the normal vector,
\be
\proj^{\ssM\ssR}N^\ssN R_{\ssM\ssN\ssR\ssT}N^\ssT=\proj^{\ssM\ssR}N^\ssN[\nabla_\ssN, \nabla_\ssR]N_\ssM\,.
\ee
The projection operator in here can be replaced by the metric, because $N^\ssM[\nabla_\ssN, \nabla_\ssR]N_\ssM=0$. This means that
\be
\proj^{\ssM\ssR}N^\ssN R_{\ssM\ssN\ssR\ssT}N^\ssT=N^\ssN[\nabla_\ssN, \nabla_\ssR]N^\ssR\,,
\ee
and the two terms in the commutator are
\ba
N^\ssN\nabla_\ssN\nabla_\ssM N^\ssM &=& N^\ssM\nabla_\ssM K\nn\\
&=&-K^2+\nabla_\ssM(KN^\ssM)\nn\\
N^\ssM\nabla_\ssN\nabla_\ssM N^\ssN &=& \nabla_\ssN\(N^\ssM\nabla_\ssM N^\ssN\) - (\nabla_\ssN N^\ssM)(\nabla_\ssM N^\ssN) \nn\\
&=&\nabla_\ssN\(N^\ssM\nabla_\ssM N^\ssN\)-K^\ssM_\ssN K^\ssN_\ssM
\ea
Combining all this we find that the curvature scalar can be written as
\be
R=\hat R + K_\ssM^\ssN K_\ssN^\ssM-K^2+2\nabla_\ssN(KN^\ssN - N^\ssM\nabla_\ssM N^\ssN)
\ee
Using this in the action, the Einstein-Hilbert term is
\ba
S_{EH} &=& -\int \exd^6x\sqrt{-g}\frac1{2\kappa^2}R\nn\\
&=& -\int \exd^6x\sqrt{-g}\frac1{2\kappa^2} \( \hat R + K_\ssM^\ssN K_\ssN^\ssM-K^2+2\nabla_\ssN(KN^\ssN - N^\ssM\nabla_\ssM N^\ssN)\)\nn\\
&=&-\int \exd^6x\sqrt{-g}\frac1{2\kappa^2} \( \hat R + K_\ssM^\ssN K_\ssN^\ssM-K^2\)-(\tilde N \cdot N)\int d^5x\sqrt{-\gamma}\frac1{2\kappa^2} 2N_\ssN\(KN^\ssN - N^\ssM\nabla_\ssM N^\ssN)\)\nn\\
&=&-\int \exd^6x\sqrt{-g}\frac1{2\kappa^2} \( \hat R + K_\ssM^\ssN K_\ssN^\ssM-K^2\)-(\tilde N \cdot N)\int d^5x\sqrt{-\gamma}\frac1{2\kappa^2} 2K\,,
\ea
Here $\tilde N$ describes the outward pointing normal, so the product $\tilde N\cdot N = \pm1$ takes care of the orientation of the boundary. The boundary integral is the negative of the Gibbons-Hawking term, as expected.

\subsection*{Boundary part of the variation}
In this section, we vary the bulk gravitational action (including the Gibbons-Hawking term) with respect to the metric. However, the only part of interest are the parts that, through stokes' theorem, have a boundary contribution. To this purpose, note that the intrinsic curvature of constant $\rho$ surfaces does not contribute (since it's derivatives are all projected into the surface). Therefore, we can take the part of the action
\ba
 S_{g,part} &=& -\int \exd^6x \sqrt{-g} \frac1{2\kappa^2} \, \left(K^\ssM_\ssN K^\ssN_\ssM-K^2\right)\nn\\
 &=&-\int \exd^6x\sqrt{-g} \frac1{2\kappa^2}
 \left(g^{\ssM\ssN}g^{\ssP\ssQ}-g^{\ssM\ssP}g^{\ssN\ssQ}\right)K_{\ssM\ssP}K_{\ssN\ssQ}\nn\\
\ea
In order to get a boundary contribution, the variation must involve a derivative. This means only the variation of the intrinsic curvatures matter:
\ba
 \delta S_{g,part} = -\int \exd^6x\sqrt{-g} \frac1{2\kappa^2}
 \left(g^{\ssM\ssN}g^{\ssP\ssQ}-g^{\ssM\ssP}g^{\ssN\ssQ}\right)2K_{\ssM\ssP} \delta K_{\ssN\ssQ}
\ea

The change in extrinsic curvature comes both from a change in normal vector, and a change in connection:
\ba
\delta(K_{\ssM\ssN})&=&\delta\left(\proj^\ssP_\ssM\proj_N^\ssQ\nabla_\ssP N_\ssQ\right)\nn\\
&=&\delta \left(\proj^\ssP_\ssM\proj_N^\ssQ\right)\nabla_\ssP N_\ssQ + \left(\proj^\ssP_\ssM\proj_N^\ssQ\right)\nabla_\ssP \delta N_\ssQ - \left(\proj^\ssP_\ssM\proj_N^\ssQ\right)(\delta\Gamma^\ssR_{\ssP\ssQ})N_\ssR
\ea
However, the variation of the projection operator has no derivative acting on it, so it is part of the bulk equations of motion. Hence,
\be
\delta S_{g,part} = -\int \exd^6x\sqrt{-g} \frac1{2\kappa^2}
 \left(g^{\ssM\ssN}g^{\ssP\ssQ}-g^{\ssM\ssP}g^{\ssN\ssQ}\right)2K_{\ssM\ssP} \proj_\ssN^\ssR \proj_\ssQ^\ssT (\nabla_\ssR \delta N_\ssT - \delta\Gamma^\ssV_{\ssR\ssT}N_\ssV )
\ee
Next, we need to find how this variation contributes as a total derivative. Stokes' theorem takes the form
\be
\int_\Omega \exd^6x \sqrt{-g}\nabla_\ssM(\lambda^\ssM) = \int_{\pd\omega}\exd^5x\sqrt{-\gamma}N_\ssM\lambda^\ssM\,,
\ee
and this shows that we can ignore any terms which have an index $R$ or $T$ on the derivative: In that case we end up contracting the normal with the projection operator, and they are perpendicular by construction (eg, $N_\ssR\proj^\ssR_\ssN=0$). The only term that satisfies this comes from the variation of the connection,
\be
\delta\Gamma^{\ssV}_{\ssR\ssT} = \frac12g^{\ssV\ssW}\(\pd_\ssR \delta g_{\ssW\ssT} + \pd_\ssT \delta g_{\ssW\ssR}-\pd_\ssW\delta g_{\ssR\ssT}\) + \cdots\,,
\ee
where the $\cdots$ indicate terms where the variation is not hit by a derivative. Clearly, the last term in this is the only one that contributes (as it is the only one that doesn't have an $R$ or $T$ derivative). This means that the variation that has a boundary contribution is
\ba
\delta S_{g,part} &=& -\int \exd^6x\sqrt{-g} \frac1{2\kappa^2}
 \left(g^{\ssM\ssN}g^{\ssP\ssQ}-g^{\ssM\ssP}g^{\ssN\ssQ}\right)2K_{\ssM\ssP} \proj_\ssN^\ssR \proj_\ssQ^\ssT \(\frac12g^{\ssV\ssW}(\pd_\ssW\delta g_{\ssR\ssT}) N_\ssV\)\nn\\
 &=&-\int \exd^6x\sqrt{-g} \frac1{2\kappa^2} \left(K^{\ssR\ssT}-K\proj^{\ssR\ssT}\right)  N^\ssW\pd_\ssW(\delta g_{\ssR\ssT})
\ea
The contribution to the boundary variation is now
\ba
\frac{\delta S}{\delta g_{\ssM\ssN}} &=& \left[\sqrt{-\gamma} \frac1{2\kappa^2} \(K^{\ssM\ssN}-K\proj^{\ssM\ssN}\)\right]_{\rho_{min}}\nn\\
 &&- \left[\sqrt{-\gamma}\frac1{2\kappa^2} \(K^{\ssM\ssN}-K\proj^{\ssM\ssN}\)\right]_{\rho_{max}} +\sum_b\frac{\delta S_b}{\delta g_{\ssM\ssN}}\,,
\ea
with $S_b$ the action on the boundary (brane). Recall that the definition of the normal and the extrinsic curvature are such that the normal is in the direction of increasing $\rho$. The contributions coming from the brane at $\rho_{min}$ are therefore
\begin{itemize}
  \item The contribution from the brane action.
  \item The 'max' contribution from the regularizing spacetime interior to the brane.
  \item The 'min' contribution from the bulk spacetime.
\end{itemize}
Demanding that the total vanishes implies
\be
\frac 1{2\kappa^2}\sqrt{-\gamma}\left[ (K^{\ssM\ssN}- K \proj^{\ssM\ssN}) - ({\rm flat})\right] = -\frac{\delta S_{-}}{\delta g_{\ssM\ssN}}\,.
\ee
Finally, we wish to make connection with the use in the main text for the codimension-2 limit. To this end, we write
\be
S_-=S_5=\frac1{2\pi}\int \exd\theta S_4\,,
\ee
which leads to the matching condition in codimension-2 from,
\be
\frac12\sqrt{g_{\theta\theta}}\sqrt{\gamma_4}\left[ (K^{\ssM\ssN}- K \proj^{\ssM\ssN}) - ({\rm flat})\right] = -\frac{\kappa^2}{2\pi}\frac{\delta S_4}{\delta g_{\ssM\ssN}}\,,
\ee
which is the form used in the main text.

\subsection*{Evaluation for the ansatz of interest}

The extrinsic curvatures for the bulk geometry of our ansatz are, at the $-\infty$ brane:
\ba
 N_\eta&=&e^{v(\eta)}(H_0\tau)^{1+c/2}\nn\\
 K_{\tau\tau}&=&\left[e^{v(\eta)}(H_0\tau)^{1+c/2}
 \right]^{-1}\omega'g_{\tau\tau}\nn\\
 K_{ij}&=&\left[e^{v(\eta)}(H_0\tau)^{1+c/2}\right]^{-1}\alpha'g_{ij}\nn\\
 K_{\theta\theta}&=&\left[e^{v(\eta)}(H_0\tau)^{1+c/2}
 \right]^{-1}\beta'g_{\theta\theta}\nn\\
 K&=&\left[e^{v(\eta)}(H_0\tau)^{1+c/2}\right]^{-1}
 (\omega'+\beta'+3\alpha')\,.
\ea
At the $+\infty$ brane, there is an overall minus sign because we need to use the oposite direction for the normal:
\ba
N_\eta&=&-e^{v(\eta)}(H_0\tau)^{1+c/2}\nn\\
K_{\tau\tau}&=&-\left[e^{v(\eta)}(H_0\tau)^{1+c/2}\right]^{-1}\omega'g_{\tau\tau}\nn\\
K_{ij}&=&-\left[e^{v(\eta)}(H_0\tau)^{1+c/2}\right]^{-1}\alpha'g_{ij}\nn\\
K_{\theta\theta}&=&-\left[e^{v(\eta)}(H_0\tau)^{1+c/2}\right]^{-1}\beta'g_{\theta\theta}\nn\\
K&=&-\left[e^{v(\eta)}(H_0\tau)^{1+c/2}\right]^{-1}(\omega'+\beta'+3\alpha')\,.
\ea

\section{Dimensional reduction}
\label{app:dimred}

This appendix performs the dimensional reduction from 6D to 4D, with the goal of relating the parameters $\kappa_4$, $f$, $C$ and $D$ to the microscopic choices for the spectator and inflaton branes. In this appendix, we again make the distinction between bare and renormalized quantities by putting an overline on the renormalized (finite) quantities.

The full 6D action is
\be
 S=S_{\rm bulk}+S_{GH}+\sum_b S_b \,,
\ee
with
\ba
 S_{\rm bulk} &=& - \int\exd^6x \sqrt{-g} \; \left\{ \frac1{2\kappa^2}
 \Bigl[ R + (\pd\phi)^2 \Bigr] + \frac14 \, e^{-\phi} \cF_{\ssM\ssN}\cF^{\ssM\ssN}
 +\frac{2g_\ssR^2}{\kappa^4}e^\phi \right\} \nn\\
 S_{\GH} &=& \sum_b \lim_{\eta\to\eta_b}
 \int \exd^5x \sqrt{-g_5} \; \left( \frac{K}{\kappa^2}
 \right) \,.
\ea
The complication is that we want to follow the zero mode, which is a combination of the metric and the scalar. In order to isolate this term, we first note that the dilaton factorizes as
\be
 \phi=\varphi_4(x^\mu)+\varphi_2(\eta,\theta)\,,
\ee
and we wish to follow $\varphi_4$ while integrating out $\varphi_2$. If we define the new metric
\be
 \tilde g_{\ssM\ssN}=g_{\ssM\ssN}e^{\varphi_4}\,,
\ee
then the extra-dimensional components are time-independent. Explicitly, we have
\be
 \tilde g_{\ssM\ssN}\exd x^\ssM\exd x^\ssN = \frac1{(H_0\tau)^2}\(-e^{2\omega}\exd\tau^2+e^{2\alpha}\delta_{ij}\exd x^i\exd x^j\) +\frac1{H_0^2}\(e^{2v}\exd\eta^2 +e^{2\beta}\exd\theta^2\)
\ee
The action under this transformation becomes
\ba
 S_{\rm bulk}&=&-\int \exd^6x\sqrt{-\tilde g}e^{-2\varphi_4}\Bigl[\frac1{2\kappa^2}\(\tilde R-4\pd_\mu\varphi_4\pd^\mu\varphi_4 +\pd_m\varphi_2\pd^m\varphi_2\)\nn\\
 &&\qquad\qquad +\frac14e^{-\varphi_2}\cF_{mn}\cF^{mn} +\frac{2g_\ssR^2}{\kappa^4}e^{\varphi_2}  \Bigr]
\ea
Since we wish to integrate out the extra-dimensinal metric, we use the trace-reversed Einstein equation (that isolates the kinetic term for the metric):
\be
 \frac1{2\kappa^2}\(\tilde g^{mn}\tilde R_{mn}+\pd_m\varphi_2\pd^m\varphi_2\)=-\frac38e^{-\varphi_2}\cF_{mn}\cF^{mn}-\frac{g_\ssR^2}{\kappa^4}e^{\varphi_2}
\ee
so we can write
\ba
 S_{\rm bulk}&=&-\int \exd^6x\sqrt{-\tilde g}e^{-2\varphi_4}\left[\frac1{2\kappa^2}\tilde g^{\mu\nu}\(\tilde R_{\mu\nu} -4\pd_\mu\varphi_4\pd_\nu\varphi_4\)-\frac18e^{-\varphi_2}\cF_{mn}\cF^{mn}+\frac{g_\ssR^2}{\kappa^4}e^{\varphi_2}\right]\nn\\
 &=&-\int \exd^6x\sqrt{-\tilde g}e^{-2\varphi_4}\left[\frac1{2\kappa^2}\tilde g^{\mu\nu}\(\tilde R_{\mu\nu} -4\pd_\mu\varphi_4\pd_\nu\varphi_4\)+\frac1{4\kappa^2}\frac1{\sqrt{\tilde g}}\pd_m\(\sqrt{-\tilde g}\pd^m\varphi_2\)\right]\,.
\ea
The next step is to separate the $\eta$ dependence of the 4-dimensional curvature, such that we can perform the integral over the extra dimensions. To this end we define
\ba
 \tilde g_{\tau\tau} &=& \hat g_{\tau\tau}(x^\mu) e^{2\omega(\eta)}\nn\\
 \tilde g_{ij} &=& \hat g_{ij}(x^\mu) e^{2\alpha(\eta)}\,,
\ea
in terms of which the 4d curvature is
\ba
 \tilde g^{\tau\tau}\tilde R_{\tau\tau}&=&e^{-2\omega}\hat g^{\tau\tau}\hat R_{\tau\tau}+H_0^2e^{-2v}\(\omega''+\omega'(\omega'+3\alpha'+\beta'-v')\)\nn\\
 \tilde g^{ij}\tilde R_{ij}&=&e^{-2\omega}\hat g^{ij}\hat R_{ij} + H_0^2e^{-2v}\(3\alpha''+3\alpha'(\omega'+3\alpha'+\beta'-v')\)
\ea
We can now split the integrals over the 4 large dimensions off in the action:
\ba
 S_{\rm bulk}&=& -\int \exd^4x \sqrt{-\hat g_4} e^{-2\varphi_4}\left[\hat R-4(\pd\varphi_4)^2\right]\int\exd^2x\frac{e^{-\omega+3\alpha+\beta+v}}{2\kappa^2H_0^2}\nn\\
 &&\qquad -\int \exd^4x\sqrt{-\hat g_4}e^{-2\varphi_4}\int \exd^2x \frac{e^{\omega+3\alpha+\beta-v}}{2\kappa^2}\(\omega''+3\alpha''+(\omega'+3\alpha')(\omega'+3\alpha'+\beta'-v')\)\nn\\
 &&\qquad -\int \exd^4x\sqrt{-\hat g_4} e^{-2\varphi_4}\int \exd^2x\frac1{4\kappa^2}\pd_m\(e^{\omega+3\alpha+\beta+v}\pd^m\varphi_2\)
\ea
The first term we can interpret the integral as a (Jordan frame) gravitational constant in 4 dimensions. The other two integrals represent the bulk contribution to the potential. Interestingly, both terms are a total derivative, leading to
\be
 S_{\rm bulk}=-\int\exd^4x\sqrt{-\hat g_4}\left\{ \frac{e^{-2\varphi_4}}{2\kappa_J^2}\left[\hat R -4(\pd\varphi_4)^2 \right] + V_{JF, \rm{bulk}}\right\}\,,
\ee
with
\be
 V_{JF, \rm{bulk}}=\frac\pi{\kappa^2}e^{-2\varphi_4}\left[e^{\omega+3\alpha+\beta-v}\(\omega'+3\alpha'+\frac{\phi'}2\)\right]_{-\infty}^\infty
\ee
Similarly we can evaluate the Gibbons-Hawking term by using that at $\eta\to\infty$ we have
\be
 K=e^{\varphi_4/2}\frac{e^{-v}}{H_0}(\omega'+3\alpha'+\beta')\,,
\ee
with oposite sign for the brane at $-\infty$. Inserting this into the Gibbons-Hawking action, and again writing it in terms of the actual dynamical variables, we find
\be
 S_{GH}=\frac{2\pi}{\kappa^2}\int \exd^4x\sqrt{-\hat g_4}\left[e^{-2\varphi_4}e^{\alpha+3\omega+\beta-v}(\omega'+3\alpha'+\beta')\right]^{\infty}_{-\infty}\,,
\ee
or
\be
 V_{JF, GH}=-\frac\pi{\kappa^2}e^{-2\varphi_4}\left[2e^{\omega+3\alpha+\beta-v}\(\omega'+3\alpha'+\beta'\)\right]_{-\infty}^\infty
\ee
Finally, we need the contribution from the branes themselves. With the same change of variables, the brane actions become
\ba
 S_s &=&-\int \exd^4x\sqrt{-\hat g_4}e^{-2\varphi_4}e^{\omega+3\alpha}\left. T_s\right|_{\eta=\infty}\nn\\
    &=&-\int \exd^4x\sqrt{-\hat g_4}e^{-2\varphi_4}e^{\cy_\infty^+-\cx_\infty^+}T_s\nn\\
 S_i &=&-\int \exd^4x\sqrt{-\hat g_4}e^{-2\varphi_4}e^{\omega+3\alpha}\left[
        T_i
       -e^{-\varphi_2}\frac{\Phi_i}2\tilde\epsilon^{mn}\cF_{mn}
       +e^{-\varphi_2}e^{-2\omega}\hat g^{\mu\nu}\pd_\mu\chi\pd_\nu\chi
       + e^{-\varphi_2}V_1e^{\zeta\chi-\varphi_4} \right]_{\eta=-\infty}\nn\\
    &=&-\int \exd^4x\sqrt{-\hat g_4}e^{-2\varphi_4} \bigg[
        e^{-\cx_\infty^-+\cy_\infty^-}\(\overline T_i-\frac{2\pi}{\kappa^2}\)
       - q H_0 \Phi_i
       +e^{-\cx_\infty^-+\cy_\infty^--\frac2c\cz_\infty^-}\hat g^{\mu\nu}\pd_\mu\overline\chi\pd_\nu\overline\chi \nn\\
       &&\qquad\qquad
       +e^{\frac32(\cy_\infty^--\cx_\infty^-)+\cz_\infty^-}\overline V_1 e^{\overline\zeta\overline\chi-\varphi_4} \bigg]
\ea
The renormailization of the tension term is a little subtle: the defining relation is
\be
 e^{(\lambda_2^--\lambda_1)\eta}\(1-\frac{\kappa^2}{2\pi}T_i\)=1-\frac{\kappa^2}{2\pi}\overline T_i\,,
\ee
and solving this for $T_i$ gives
\be \label{eq:Trenorm}
 e^{(\lambda_2^--\lambda_1)\eta}T_i=\overline T_i+\frac{2\pi}{\kappa^2}\(e^{(\lambda_2^--\lambda_1)\eta}-1\)\,.
\ee
If $\lambda_2^-=\lambda_1$, the second term vanishes. However, for the inflationary solutions we have $\lambda_3^-\neq0$, which through the constraint ~\pref{eq:powersconstraint} implies $\lambda_2^->\lambda_1$ for finite volume solutions. This means that the exponent in ~\pref{eq:Trenorm} vanishes at $\eta\to-\infty$, leaving the constant part.

\subsubsection*{Relating the bulk and GH contributions to the branes}

Adding the bulk and GH terms we need to evaluate
\be
S_B\supset-\int \exd^4x\sqrt{\hat g_4}\frac\pi{\kappa^2}e^{-2\varphi_4}\left[e^{\omega+3\alpha+\beta-v}\(-\omega'-3\alpha'-2\beta'+\frac12\varphi_2'\)\right]_{-\infty}^{\infty}\,.
\ee
Evaluating this in terms of the solution functions, $\cx$, $\cy$ and $\cz$ shows that the exponential is identically $1$. Replacing the derivatives by asymptotic powers $\lambda_i$, this becomes
\ba
S_B&\supset&-\int \exd^4x\sqrt{\hat g_4}\frac\pi{\kappa^2}e^{-2\varphi_4}\left[ -\frac14\lambda_1-\frac74\lambda_2^\pm-\frac{3}{2}\lambda_3^\pm \right]_{-\infty}^{\infty}\nn\\
&=&-\int \exd^4x\sqrt{\hat g_4}\frac\pi{\kappa^2}e^{-2\varphi_4}\left[\frac14(\lambda_1+\lambda_1)+\frac74(\lambda_2^-+\lambda_2^+) + \frac{3}{2}(\lambda_3^-+\lambda_3^+)\right]\nn\\
&=&-\int \exd^4x\sqrt{\hat g_4}\frac\pi{\kappa^2}e^{-2\varphi_4}\left[\frac12\lambda_1+\frac74(\lambda_2^--\lambda_1) + \frac{3}{2}\lambda_3^- \right]\nn\\
&=&-\int \exd^4x\sqrt{\hat g_4}\frac\pi{\kappa^2}e^{-2\varphi_4}\left[\frac94\lambda_1+\frac74\lambda_2^-+\frac{3}{2}\lambda_3^-\right]
\ea
where we have used the matching conditions at the spectator ($\eta\to\infty$) brane, $\lambda_3^+=0$ and $\lambda_2^+=\lambda_1$.

We can rewrite this combination of powers in terms of the combination
\be
\frac94\lambda_1+\frac74\lambda_2^-+\frac32\lambda_3^-= \frac92\phi' +\frac{24+9c}{7+2c}\left[\beta'+2\alpha'+\omega'\right] +\frac{4-c}{7+2c}\left[\beta'+3\alpha'\right]\,,
\ee
which we can relate to derivatives of the brane action:
\ba
\phi'&=&e^{v-\beta}\frac{\kappa^2}{2\pi}\frac1{\sqrt{-\gamma}}\frac{\delta S_i}{\delta\phi}\nn\\
\beta'+3\alpha'&=&e^{v-\beta}\(1+\frac{\kappa^2}{\pi}\frac1{\sqrt{-\gamma}}\frac{\delta S_i}{\delta g_{\tau\tau}}g_{\tau\tau}\)\nn\\
\beta'+2\alpha'+\omega'&=&e^{v-\beta}\(1+\frac13\frac{\kappa^2}{\pi}\frac1{\sqrt{-\gamma}}\frac{\delta S_i}{\delta g_{ij}}g_{ij}\)
\ea
Evaluating this in terms of the low-energy modes we get
\ba
\phi'&=&\frac{\kappa^2}{2\pi}\left[ -H_0q\Phi_i+ e^{-\cx_\infty^-+\cy_\infty^--\frac2c\cz_\infty^-}\hat g^{\mu\nu}\pd_\mu\overline\chi\pd_\nu\overline\chi + e^{-\frac32\cx_\infty^-+\frac32\cy_\infty^-+\cz_\infty^-}\overline V_1 e^{\overline\zeta\overline\chi-\varphi_4}  \right]\nn\\
(\beta+3\alpha)'&=&e^{-\cx_\infty^-+\cy_\infty^-}\(1-\frac{\kappa^2}{2\pi}\overline T_i\)+\frac{\kappa^2}{2\pi}\bigg(H_0q\Phi_i + e^{-\cx_\infty^-+\cy_\infty^--\frac2c\cz_\infty^-}\hat g^{\mu\nu}\pd_\mu\overline\chi\pd_\nu\overline\chi\nn\\
&&\qquad\qquad\qquad\qquad - e^{-\frac32\cx_\infty^-+\frac32\cy_\infty^-+\cz_\infty^-}\overline V_1 e^{\overline\zeta\overline\chi-\varphi_4} \bigg) \nn\\
(\beta+2\alpha+\omega')&=&e^{-\cx_\infty^-+\cy_\infty^-}\(1-\frac{\kappa^2}{2\pi}\overline T_i\)+\frac{\kappa^2}{2\pi}\bigg(H_0q\Phi_i - e^{-\cx_\infty^-+\cy_\infty^--\frac2c\cz_\infty^-}\hat g^{\mu\nu}\pd_\mu\overline\chi\pd_\nu\overline\chi\nn\\
&&\qquad\qquad\qquad\qquad - e^{-\frac32\cx_\infty^-+\frac32\cy_\infty^-+\cz_\infty^-}\overline V_1 e^{\overline\zeta\overline\chi-\varphi_4} \bigg) \nn\\
\ea

\subsection*{The total action}

The kinetic parts of the action take the form
\be
S_k=-\int \exd^4x\sqrt{-\hat g_4}e^{-2\varphi_4}\left[\frac1{2\kappa_{JF}^2}\hat g^{\mu\nu}(\hat R_{\mu\nu}-4\pd_\mu\varphi_4\pd_\nu\varphi_4) + f^2\hat g^{\mu\nu}\pd_\mu\overline\chi\pd_\nu\overline\chi \right]\,.
\ee
where the lines on the field $\chi$ represent the required wavefunction renormalization. We introduce a decay constant $f$ for the inflaton, because we are only guaranteed that the coefficient of the kinetic term is rendered finite by renormalization, not that it is unity in the effective theory. Explicitly we find
\be
f^{2}=e^{-\cx_\infty^-+\cy_\infty^--\frac2c\cz_\infty^-}\left(\frac{23-2c}{28+8c}\right)
\ee
As for the potential, there are two types of terms:
\be
S_{V}=-\int \exd^4x\sqrt{-\hat g_4}\left[Ce^{-2\varphi_4} +De^{\overline\zeta\overline\chi-3\varphi_4} \right]\,.
\ee
where we can again explicitly evaluate the constants as
\ba
C&=& e^{-\cx_\infty^++\cy_\infty^+}\overline T_s - \frac54 qH_0\Phi_i + e^{-\cx_\infty^-+\cy_\infty^-}\(\frac{2\pi}{\kappa^2}-\overline T_i\) \nn\\
D&=&\frac54e^{-\frac32\cx_\infty^-+\frac32\cy_\infty^-+\cz_\infty^-}\overline V_1\,.
\ea
Finally we need to go to the 4d Einstein frame, for which we have
\be
{\bf g}_{\mu\nu}= e^{-2\varphi_4}\hat g_{\mu\nu}
\ee
and the potential becomes
\be
V_{EF}= Ce^{2\varphi_4} + De^{\overline\zeta\overline\chi+\varphi_4}\,.
\ee
The kinetic term after the conformal transformation becomes
\be
S_k=-\int \exd^4x\sqrt{-{\bf g}_4}\left[\frac1{2\kappa_{JF}^2}{\bf g}^{\mu\nu}({\bf R}_{\mu\nu}+2\pd_\mu\varphi_4\pd_\nu\varphi_4) + f^2{\bf g}^{\mu\nu}\pd_\mu\overline\chi\pd_\nu\overline\chi \right]\,.
\ee


\begin{thebibliography}{999}

\bibitem{MirageCosmology}
 A. Kehagias, E. Kiritsis,
 ``Mirage Cosmology,'' JHEP {\bf 9911} (1999) 022
 [arXiv:hep-th/9910174];

\bibitem{RScosmo}
 Horace Stoica, S.-H. Henry Tye, Ira Wasserman,
 ``Cosmology in the Randall-Sundrum Brane World Scenario''
 Phys.\ Lett.\ {\bf B482} (2000) 205-212;

 Peter Bowcock, Christos Charmousis, Ruth Gregory,
 ``General brane cosmologies and their global spacetime structure''
 Class. \ Quant. \ Grav. \  {\bf 17} (2000) 4745-4764;

   A.~Hebecker, J.~March-Russell,
  ``Randall-Sundrum II cosmology, AdS / CFT, and the bulk black hole,''
  Nucl.\ Phys.\  {\bf B608 } (2001)  375-393.
  [hep-ph/0103214];

   D.~Langlois,
  ``Brane cosmology: An Introduction,''
  Prog.\ Theor.\ Phys.\ Suppl.\  {\bf 148 } (2003)  181-212.
  [hep-th/0209261];

  R.~Maartens,
  ``Brane world gravity,''
  Living Rev.\ Rel.\  {\bf 7 } (2004)  7,
  [gr-qc/0312059];

   C.~Csaki, M.~Graesser, L.~Randall, J.~Terning,
  ``Cosmology of brane models with radion stabilization,''
  Phys.\ Rev.\  {\bf D62 } (2000)  045015.
  [hep-ph/9911406];

\bibitem{IJC}
   K.~Lanczos, Phys.\ Z.\ {\bf 23} (1922) 239--543; Ann.\ Phys.\
  {\bf 74} (1924) 518--540;

  C.W.~Misner and D.H.~Sharp,
  ``Relativistic Equations for Adiabatic, Spherically Symmetric
  Gravitational Collapse''
  Phys.\ Rev.\ {\bf 136} (1964)
  571--576;

  W.~Israel,
  ``Singular hypersurfaces and thin shells in general relativity''
  Nuov.\ Cim.\ {\bf 44B} (1966) 1--14; {\it errata}
  Nuov.\ Cim.\ {\bf 48B} 463.

\bibitem{SIreviews}
    F.~Quevedo,
    ``Lectures on String/Brane Cosmology,''
    Class.\ Quant.\ Grav.\  {\bf 19} (2002) 5721
    [arXiv:hep-th/0210292];

    E.~Kiritsis,
  ``D-branes in standard model building, gravity and cosmology,''
  Phys.\ Rept.\  {\bf 421 } (2005)  105-190.
  [hep-th/0310001];

   A.~Linde,
  ``Inflation and string cosmology,''
  eConf {\bf C040802} (2004) L024
  [J.\ Phys.\ Conf.\ Ser.\  {\bf 24} (2005) 151]
  [arXiv:hep-th/0503195];

  S.~-H.~Henry Tye,
  ``Brane inflation: String theory viewed from the cosmos,''
  Lect.\ Notes Phys.\  {\bf 737 } 949 (2008)
  [arXiv:hep-th/0610221];

   R.~Kallosh,
  ``On inflation in string theory,''
  Lect.\ Notes Phys.\  {\bf 738 } 119 (2008)
  [arXiv:hep-th/0702059];

   C.~P.~Burgess,
  ``Lectures on Cosmic Inflation and its Potential Stringy Realizations,''
  Class.\ Quant.\ Grav.\  {\bf 24 }  S795 (2007)
  [arXiv:0708.2865 [hep-th]];

   L.~McAllister and E.~Silverstein,
  ``String Cosmology: A Review,''
  Gen.\ Rel.\ Grav.\  {\bf 40 } 565 (2008)
  [arXiv:0710.2951 [hep-th]].

\bibitem{scaling solutions}
   A.~J.~Tolley, C.~P.~Burgess, C.~de Rham, D.~Hoover,
  ``Scaling solutions to 6D gauged chiral supergravity,''
  New J.\ Phys.\  {\bf 8 } (2006)  324.
  [hep-th/0608083].

\bibitem{NS}
 H. Nishino and E. Sezgin, {\it Phys. Lett.} {\bf 144B} (1984) 187;
 ``The Complete N=2, D = 6 Supergravity With Matter And Yang-Mills
 Couplings,'' Nucl.\ Phys.\ {\bf B278} (1986) 353;

 S. Randjbar-Daemi, A. Salam, E. Sezgin and J. Strathdee,
 ``An Anomaly Free Model in Six-Dimensions''
 {\it
 Phys. Lett.} {\bf B151} (1985) 351.



\bibitem{HML}
  H.~M.~Lee, A.~Papazoglou,
  ``Codimension-2 brane inflation,''
  Phys.\ Rev.\  {\bf D80 } (2009)  043506.
  [arXiv:0901.4962 [hep-th]];

  J.~Gallicchio, R.~Mahbubani,
  ``Inflation on the Brane with Vanishing Gravity,''
  JHEP {\bf 1004 } (2010)  068.
  [arXiv:0911.5343 [hep-th]].


\bibitem{GW}
     W.~D.~Goldberger and M.~B.~Wise,
     ``Modulus stabilization with bulk fields'',
     Phys.\ Rev.\ Lett.\ {\bf 83} (1999) 4922-4925
     [arXiv:hep-ph/9907447].

\bibitem{Cod2Matching}
  C.~P.~Burgess, D.~Hoover, G.~Tasinato,
  ``UV Caps and Modulus Stabilization for 6D Gauged Chiral Supergravity,''
  JHEP {\bf 0709 } (2007)  124.
  [arXiv:0705.3212 [hep-th]];

  C.~P.~Burgess, D.~Hoover, C.~de Rham, G.~Tasinato,
  ``Effective Field Theories and Matching for Codimension-2 Branes,''
  JHEP {\bf 0903 } (2009)  124.
  [arXiv:0812.3820 [hep-th]].


\bibitem{BBvN}
   A.~Bayntun, C.~P.~Burgess, L.~van Nierop,
  ``Codimension-2 Brane-Bulk Matching: Examples from Six and Ten Dimensions,''
  New J.\ Phys.\  {\bf 12 } (2010)  075015.
  [arXiv:0912.3039 [hep-th]].


\bibitem{BulkAxions}
 C.~P.~Burgess, L.~van Nierop,
  ``Bulk Axions, Brane Back-reaction and Fluxes,''
  JHEP {\bf 1102 } (2011)  094.
  [arXiv:1012.2638 [hep-th]].

\bibitem{susybranes}
  C.~P.~Burgess, L.~van Nierop,
  ``Large Dimensions and Small Curvatures from Supersymmetric Brane Back-reaction,''
  JHEP {\bf 1104 } (2011)  078.
  [arXiv:1101.0152 [hep-th]].


\bibitem{6DdS}
 A.~J.~Tolley, C.~P.~Burgess, D.~Hoover, Y.~Aghababaie,
  ``Bulk singularities and the effective cosmological constant for higher co-dimension branes,''
  JHEP {\bf 0603 } (2006)  091.
  [hep-th/0512218].


\bibitem{ExtInf}
  D.~La, P.~J.~Steinhardt,
  ``Extended Inflationary Cosmology,''
  Phys.\ Rev.\ Lett.\  {\bf 62 } (1989)  376;
  %
  ``Bubble Percolation in Extended Inflationary Models,''
  Phys.\ Lett.\  {\bf B220 } (1989)  375;

   D.~La, P.~J.~Steinhardt, E.~W.~Bertschinger,
  ``Prescription for Successful Extended Inflation,''
  Phys.\ Lett.\  {\bf B231 } (1989)  231;


\bibitem{SSs}
 A.~Salam and E.~Sezgin,
 ``Chiral Compactification On Minkowski X S**2 Of N=2
 Einstein-Maxwell Supergravity In Six-Dimensions,''
 Phys.\ Lett.\ B {\bf 147} (1984) 47.

\bibitem{ConTrunc}
    G.~W.~Gibbons, C.~N.~Pope,
  ``Consistent S**2 Pauli reduction of six-dimensional chiral gauged Einstein-Maxwell supergravity,''
  Nucl.\ Phys.\  {\bf B697 } (2004)  225-242.
  [hep-th/0307052].

\bibitem{CopelandSeto}
 Edmund J. Copeland, Osamu Seto,
 ``Dynamical solutions of warped six dimensional supergravity''
 JHEP {\bf 0708} (2007) 001
 [arXiv:0705.4169 [hep-th]]

\bibitem{dSnogo}
  J. Maldacena and C. Nunez,
 ``Supergravity description of field theories on curved manifolds and a no go theorem,'' Int. J. Mod. Phys. {\bf A16} (2001) 822-855
 [arXiv:hep-th/0007018];

   D.~H.~Wesley,
  ``Oxidised cosmic acceleration,''
  JCAP {\bf 0901 } (2009)  041,
  [arXiv:0802.3214 [hep-th]];

 P.J. Steinhardt and D. Wesley,
  ``Dark Energy, Inflation and Extra Dimensions,''
 Phys. Rev. {\bf D79} 104026 (2009);

\bibitem{TNCC}
   Y.~Aghababaie, C.~P.~Burgess, S.~L.~Parameswaran, F.~Quevedo,
  ``Towards a naturally small cosmological constant from branes in 6-D supergravity,''
  Nucl.\ Phys.\  {\bf B680 } (2004)  389-414
  [hep-th/0304256];

   C.~P.~Burgess,
  ``Towards a natural theory of dark energy: Supersymmetric large extra dimensions,''
  AIP Conf.\ Proc.\  {\bf 743 } (2005)  417-449
  [hep-th/0411140];
  %
  ``Supersymmetric large extra dimensions and the cosmological constant: An Update,''
  Annals Phys.\  {\bf 313 } (2004)  283-401
  [hep-th/0402200];

  C.~P.~Burgess, L.~van Nierop,
  ``Technically Natural Cosmological Constant From Supersymmetric 6D Brane Backreaction,''
  [arXiv:1108.0345 [hep-th]].


\bibitem{RS}
  L. Randall, R. Sundrum,
  ''A Large Mass Hierarchy from a Small Extra Dimension''
  { Phys.\ Rev.\ Lett.} {\bf 83}
  (1999) 3370 [hep-ph/9905221];
  Phys.\ Rev.\ Lett.\ {\bf 83} (1999)
  4690 [hep-th/9906064].

\bibitem{PST}
  M.~Peloso, L.~Sorbo, G.~Tasinato,
  ``Standard 4-D gravity on a brane in six dimensional flux compactifications,''
  Phys.\ Rev.\  {\bf D73 } (2006)  104025.
  [hep-th/0603026];

   B.~Himmetoglu and M.~Peloso,
  ``Isolated Minkowski vacua, and stability analysis for an extended brane in
  the rugby ball,''
  Nucl.\ Phys.\  B {\bf 773} (2007) 84
  [hep-th/0612140].

\bibitem{otheruvcaps}
 P.~Bostock, R.~Gregory, I.~Navarro and J.~Santiago,
  ``Einstein gravity on the codimension 2 brane?,''
  Phys.\ Rev.\ Lett.\  {\bf 92} (2004) 221601
  [arXiv:hep-th/0311074];

   I.~Navarro and J.~Santiago,
  ``Gravity on codimension 2 brane worlds,''
  JHEP {\bf 0502} (2005) 007
  [arXiv:hep-th/0411250];

  J.~Vinet and J.~M.~Cline,
  ``Codimension-two branes in six-dimensional supergravity and the
  cosmological constant problem,''
  Phys.\ Rev.\  D {\bf 71} (2005) 064011
  [hep-th/0501098];

   C.~de Rham,
  ``The Effective Field Theory of Codimension-two Branes,''
  JHEP {\bf 0801} (2008) 060
  [arXiv:0707.0884 [hep-th]];

  E.~Papantonopoulos, A.~Papazoglou and V.~Zamarias,
  ``Regularization of conical singularities in warped six-dimensional
  compactifications,''
  JHEP {\bf 0703} (2007) 002
  [arXiv:hep-th/0611311];
%
  ``Induced cosmology on a regularized brane in six-dimensional flux
  compactification,''
  Nucl.\ Phys.\  B {\bf 797} (2008) 520
  [arXiv:0707.1396 [hep-th]];

     D.~Yamauchi and M.~Sasaki,
  ``Brane World in Arbitrary Dimensions Without $Z_2$ Symmetry,''
  Prog.\ Theor.\ Phys.\  {\bf 118} (2007) 245
  [arXiv:0705.2443 [gr-qc]];

   N.~Kaloper and D.~Kiley,
  ``Charting the Landscape of Modified Gravity,''
  JHEP {\bf 0705} (2007) 045
  [hep-th/0703190];

  M.~Minamitsuji and D.~Langlois,
  ``Cosmological evolution of regularized branes in 6D warped flux
  compactifications,''
  Phys.\ Rev.\  D {\bf 76} (2007) 084031
  [arXiv:0707.1426 [hep-th]];

   S.~A.~Appleby and R.~A.~Battye,
  ``Regularized braneworlds of arbitrary codimension,''
  Phys.\ Rev.\  D {\bf 76} (2007) 124009
  [arXiv:0707.4238 [hep-ph]];

  C.~Bogdanos, A.~Kehagias and K.~Tamvakis,
  ``Pseudo-3-Branes in a Curved 6D Bulk,''
  Phys.\ Lett.\  B {\bf 656} (2007) 112
  [arXiv:0709.0873 [hep-th]];

 O.~Corradini, K.~Koyama and G.~Tasinato,
  ``Induced gravity on intersecting brane-worlds Part I: Maximally symmetric
  solutions,''
  Phys.\ Rev.\  D {\bf 77} (2008) 084006
  [arXiv:0712.0385 [hep-th]];
  ``Induced gravity on intersecting brane-worlds Part II: Cosmology,''
  Phys.\ Rev.\  D {\bf 78} (2008) 124002
  [arXiv:0803.1850 [hep-th]];

    F.~Arroja, T.~Kobayashi, K.~Koyama and T.~Shiromizu,
  ``Low energy effective theory on a regularized brane in 6D gauged chiral
  supergravity,''
  JCAP {\bf 0712} (2007) 006
  [arXiv:0710.2539 [hep-th]];

   V.~Dzhunushaliev, V.~Folomeev and M.~Minamitsuji,
  ``Thick brane solutions,''
  arXiv:0904.1775 [gr-qc].


\bibitem{TvsA}
 A. Vilenkin,
 ``Gravitational field of vacuum domain walls and strings''
 Phys. Rev. {\bf D23} (1981) 852;

 R. Gregory,
 ``Gravitational stability of local strings''
 Phys. Rev. Lett. {\bf 59} (1987) 740;

 A.~G.~Cohen and D.~B.~Kaplan,
 ``The Exact Metric About Global Cosmic Strings,''
 {\it Phys.\ Lett.} {\bf B215}, 67 (1988);

 A. Vilenkin and P. Shellard, {\em Cosmic Strings and other
 Topological Defects}, Cambridge University Press (1994);

 R.~Gregory and C.~Santos,
 ``Cosmic strings in dilaton gravity,''
 Phys.\ Rev.\ D {\bf 56}, 1194 (1997) [gr-qc/9701014].


\bibitem{Bren}
 W.~D.~Goldberger, M.~B.~Wise,
  ``Renormalization group flows for brane couplings,''
  Phys.\ Rev.\  {\bf D65 } (2002)  025011
  [hep-th/0104170];

   E.~Dudas, C.~Papineau, V.A.~Rubakov,
   ``Flowing to four dimensions''
   JHEP0603:085,2006
   arXiv:hep-th/0512276v1;

   C.~de Rham,
  ``The Effective Field Theory of Codimension-two Branes,''
  JHEP {\bf 0801} (2008) 060
  [arXiv:0707.0884 [hep-th]];

 C.~P.~Burgess, C.~de Rham, L.~van Nierop,
  ``The Hierarchy Problem and the Self-Localized Higgs,''
  JHEP {\bf 0808 } (2008)  061.
  [arXiv:0802.4221 [hep-ph]].

\bibitem{ExtInfKK}
    R.~Holman, E.~W.~Kolb, S.~L.~Vadas, Y.~Wang,
  ``Extended inflation from higher dimensional theories,''
  Phys.\ Rev.\  {\bf D43 } (1991)  995-1004.

\bibitem{ExtInfStr}
    B.~A.~Campbell, A.~D.~Linde, K.~A.~Olive,
  ``Does string theory lead to extended inflation?,''
  Nucl.\ Phys.\  {\bf B355 } (1991)  146-161.

\bibitem{ExtInfProb}
   E.~J.~Weinberg,
  ``Some Problems with Extended Inflation,''
  Phys.\ Rev.\  {\bf D40 } (1989)  3950;

    D.~S.~Salopek, J.~R.~Bond, J.~M.~Bardeen,
  ``Designing Density Fluctuation Spectra in Inflation,''
  Phys.\ Rev.\  {\bf D40 } (1989)  1753;

   A.~D.~Linde,
  ``Eternal extended inflation and graceful exit from old inflation without Jordan-Brans-Dicke,''
  Phys.\ Lett.\  {\bf B249 } (1990)  18-26;

   E.~W.~Kolb, D.~S.~Salopek, M.~S.~Turner,
  ``Origin of density fluctuations in extended inflation,''
  Phys.\ Rev.\  {\bf D42 } (1990)  3925-3935;

   R.~Fakir, W.~G.~Unruh,
  ``Improvement on cosmological chaotic inflation through nonminimal coupling,''
  Phys.\ Rev.\  {\bf D41 } (1990)  1783-1791;

  R.~Holman, E.~W.~Kolb, Y.~Wang,
  ``Gravitational Couplings Of The Inflaton In Extended Inflation,''
  Phys.\ Rev.\ Lett.\  {\bf 65 } (1990)  17-20;

   J.~Garcia-Bellido, M.~Quiros,
  ``Extended Inflation In Scalar - Tensor Theories Of Gravity,''
  Phys.\ Lett.\  {\bf B243 } (1990)  45-51;

\bibitem{VolInf}
  J.~P.~Conlon, R.~Kallosh, A.~D.~Linde, F.~Quevedo,
  ``Volume Modulus Inflation and the Gravitino Mass Problem,''
  JCAP {\bf 0809 } (2008)  011.
  [arXiv:0806.0809 [hep-th]].


\bibitem{6Dquint}
 A.~Albrecht, C.~P.~Burgess, F.~Ravndal, C.~Skordis,
  ``Natural quintessence and large extra dimensions,''
  Phys.\ Rev.\  {\bf D65 } (2002)  123507.
  [astro-ph/0107573].

\bibitem{ADD}
  N.~Arkani-Hamed, S.~Dimopoulos and G.~R.~Dvali,
  ``The hierarchy problem and new dimensions at a millimeter,''
  Phys.\ Lett.\ B {\bf 429} (1998) 263
  [arXiv:hep-ph/9803315];

  I.~Antoniadis, N.~Arkani-Hamed, S.~Dimopoulos and G.~R.~Dvali,
  ``New dimensions at a millimeter to a Fermi and superstrings at a TeV,''
  Phys.\ Lett.\ B {\bf 436} (1998) 257
  [arXiv:hep-ph/9804398].

\bibitem{MSLED}
 C.~P.~Burgess, J.~Matias, F.~Quevedo,
  ``MSLED: A Minimal supersymmetric large extra dimensions scenario,''
  Nucl.\ Phys.\  {\bf B706 } (2005)  71-99.
  [hep-ph/0404135].

\bibitem{QInf}
   P.~J.~E.~Peebles, A.~Vilenkin,
  ``Quintessential inflation,''
  Phys.\ Rev.\  {\bf D59 } (1999)  063505.
  [astro-ph/9810509];

  V.~Faraoni,
  ``Inflation and quintessence with nonminimal coupling,''
  Phys.\ Rev.\  {\bf D62 } (2000)  023504.
  [gr-qc/0002091];

  G.~Huey, J.~E.~Lidsey,
  ``Inflation, brane worlds and quintessence,''
  Phys.\ Lett.\  {\bf B514 } (2001)  217-225.
  [astro-ph/0104006];

   A.~S.~Majumdar,
  ``From brane assisted inflation to quintessence through a single scalar field,''
  Phys.\ Rev.\  {\bf D64 } (2001)  083503.
  [astro-ph/0105518];

    M.~Malquarti, A.~RLiddle,
  ``Initial conditions for quintessence after inflation,''
  Phys.\ Rev.\  {\bf D66 } (2002)  023524.
  [astro-ph/0203232];

   P.~Brax, J.~Martin,
  ``Coupling quintessence to inflation in supergravity,''
  Phys.\ Rev.\  {\bf D71 } (2005)  063530.
  [astro-ph/0502069];

    M.~Malquarti, A.~RLiddle,
  ``Initial conditions for quintessence after inflation,''
  Phys.\ Rev.\  {\bf D66 } (2002)  023524.
  [astro-ph/0203232].

\bibitem{DERev}
 E.~J.~Copeland, M.~Sami, S.~Tsujikawa,
  ``Dynamics of dark energy,''
  Int.\ J.\ Mod.\ Phys.\  {\bf D15 } (2006)  1753-1936.
  [hep-th/0603057].

\bibitem{CP}
  M.~Cvetic, G.~W.~Gibbons, C.~N.~Pope,
  ``A String and M theory origin for the Salam-Sezgin model,''  Nucl.\ Phys.\  {\bf B677 } (2004)  164-180.
  [hep-th/0308026];

   T.~G.~Pugh, E.~Sezgin, K.~S.~Stelle,
  ``D=7 / D=6 Heterotic Supergravity with Gauged R-Symmetry,''
  JHEP {\bf 1102 } (2011)  115.
  [arXiv:1008.0726 [hep-th]].


\end{thebibliography}
\end{document}